\documentclass[preprint,12pt]{elsarticle}
\usepackage{setspace}
\usepackage{amsthm}
\theoremstyle{definition}

\usepackage{url}
\usepackage{graphicx}
\usepackage{dcolumn}
\usepackage{multirow}
\usepackage{subfigure}
\usepackage{times,mathptm}
\usepackage{float}
\usepackage{color}
\usepackage{amsmath,amsfonts}
\usepackage{mathptmx}
\usepackage{mathrsfs}
\usepackage{bbm}
\usepackage{bm}
\usepackage{xfrac}

\newcommand{\beq}{\begin{equation}}
\newcommand{\eeq}{\end{equation}} 
\newcommand{\bea}{\begin{eqnarray}}
\newcommand{\eea}{\end{eqnarray}}

\renewcommand{\d}{\delta}

\renewcommand{\l}{\lambda}

\newcommand{\T}{{\cal T}}

\renewcommand{\l}{\lambda}

\renewcommand{\a}{\alpha}

\renewcommand{\ni}{\noindent}
\newcommand{\tr}{\text{Tr}}

\newcommand{\bx}{\mathbf{x}}
\newcommand{\by}{\mathbf{y}}

\newcommand{\e}{\epsilon}

\renewcommand{\k}{\kappa}
\newcommand{\D}{\Delta}

\newcommand{\oh}{\frac{1}{2}}

\newcommand{\non}{\nonumber}

\newcommand{\rf}[1]{(\ref{#1})}
\newcommand{\ra}{\rightarrow}

\usepackage{empheq}

\usepackage{ulem}

\begin{document}

\begin{frontmatter}
\title{Loss of ergodicity in a quantum hopping model of \\ 
 a dense many body system with repulsive interactions}

\author{Kazue Matsuyama}
\ead{kazuem@sfsu.edu}
\address{Physics and Astronomy Department \\ San Francisco State
University  San Francisco, CA~94132, USA}
\begin{abstract}
In this work we report on a loss of ergodicity in a simple hopping model, motivated by the Hubbard Hamiltonian, of a many body quantum system at zero temperature, quantized in Euclidean time.  We show that this quantum system may lose ergodicity at high densities on a large lattice, as a result of both Pauli exclusion and strong Coulomb repulsion. In particular we study particle hopping susceptibilities and the tendency towards particle localization.  It is found that the appearance and existence of quantum phase transitions in this model, in the case of high density and strong Coulomb repulsion, depends on the starting configuration of particle trajectories in the numerical simulation. We argue that this breakdown may be the Euclidean time version of a breakdown of the eigenstate thermalization hypothesis in real time quantization.   
\end{abstract}


\end{frontmatter}

\bibliographystyle{elsarticle-num} 

\section{\label{sec:level1}Introduction}
In recent years there has been some interest in the phenomena of non-ergodicity and very slow thermalization
in quantum many-body systems \cite{garrahan} associated with many-body localization in random potentials \cite{Abanin2019,Altman2015,Nand2015}, and 
with certain translation-invariant quantum hopping models \cite{lan2018quantum,van2015dynamics}.
There has also been a great deal of effort devoted to the sign problem in the Hubbard model at finite density, although this is still a work in progress. 
Motivated largely by the sign problem in the Hubbard model, I investigate here the Euclidean-time quantization, 
in discretized time, of a many-body hopping model incorporating features which are reminiscent of the Hubbard model.  
What I will show here is that this quantized hopping model, which can also be regarded as the statistical mechanics of 
interacting particle trajectories, exhibits clear non-ergodic behavior, i.e.\ a strong and qualitative dependence of 
expectation values on the initial configuration, which includes the appearance, or non-appearance, of quantum phase
transitions depending on the initial state.  This seems to be another example, in addition to 
the cases cited above, of non-ergodicity in a translation-invariant quantum system, in which 
there is presumably a violation of the eigenstate thermalization hypothesis \cite{Deutsch1991,Srednicki1994,Tasaki1998,Rigol2008}.

\section{\label{sec:level1}The model}

  The initial motivation was to approximate some features of the Hubbard Hamiltonian
\beq
  H  = - t\sum_{\langle j,i\rangle,{\sigma}}(c^{\dagger}_{j\sigma}c_{i\sigma} + c^{\dagger}_{i\sigma}c_{j\sigma}) + U\sum_j n_{j\uparrow}n_{j\downarrow} - \mu\sum_j(n_{j\uparrow} + n_{j\downarrow})
  \label{HH}
\eeq  
in a simpler model which is hopefully more tractable to numerical simulation at high densities. It is a long-standing
conjecture, in the condensed matter community, that this simple Hamiltonian system describes the behavior of
strongly correlated electrons in solids.  The roadblock, at least so far as standard numerical algorithms are concerned, is the
sign problem.  There are a number of approaches, e.g.\ the complex Langevin equation \cite{Aarts:2009uq}, deformation of path integration into the complex plane (the ``thimble'' approach) \cite{Cristoforetti:2013wha}, and the density of states method \cite{Langfeld:2012ah}, which have been
investigated, largely in the high energy physics community, in an effort to deal with the sign problem associated with the QCD phase diagram.  Although successes with these methods are so far limited, there have been recent efforts to import them to deal with the sign problem in many body systems \cite{Berger:2019odf,Korner:2020vjw,Fukuma:2019wbv,Mukherjee:2014hsa}.  Whether one or more of these approaches will be useful in the case of the Hubbard model (or the QCD phase diagram, for that matter) is not yet clear; what is known is that the problem of  finding a {\it general} solution of the sign problem lies in the NP-hard complexity class \cite{Troyer:2004ge}.

    In this article we do not attempt a direct simulation of the Hubbard model.  Instead we construct a simplified
hopping model which is free of the sign problem, but which might incorporate at least some of the same physics.  The model
contains two types of particles, which we refer to as ``spin up'' and ``spin down,'' with a property which we
will refer to, a little loosely, as the ``Exclusion Principle," meaning that  no two particles of the same type can occupy the same lattice site.   Euclidean time is also discretized, and a particle may hop in one time step to either a nearest or next-nearest lattice site if (i) this transition is allowed by the Exclusion Principle; and (ii) the spatial separation between a particle's position at time step $t$, and the position at time steps $t \pm 1$, does not exceed the nearest and next-nearest criterion.  The Euclidean action is given by
\bea
        S &=& \sum_{t=1}^{N_t} \left\{\k  \sum_{n=1}^{n_p} j(n,t) + \sum_{x,y} V(x,y,t) \right\} \non \\
            &=& \sum_{n=1}^{n_p} K(n) + \sum_{x,y,t} V(x,y,t)
   \label{action}
\eea
where $n_p$ is the total number of particles on the finite lattice, $N_t$ is the extension in the time direction, with
\beq
  j(n,t)=\begin{cases}
               0 ~~ \text{if particle $n$ is at the same lattice site at} \\
               ~~~~~\text{time~} t+1\\
               1 ~~\text{if particle $n$ is at a nearest or next-nearest }  \\
               ~~~~~\text{site at time~} t+1
            \end{cases} \label{j}
\eeq
as compared to the site occupied by particle $n$ at time $t$, and
\beq
  V(x, y, t)=\begin{cases}
               0 ~~ \text{if zero or one electron at site~} x,y\\
               U ~~\text{if two electrons of opposite spins at site~} x,y\\
            \end{cases}
            \label{pot}
\eeq 
at time $t$.
We have $K(n)=\k ~\times$ the number of hops along the trajectory of the $n$-th particle, and we count it as one ``hop'' whenever
${j(n,t)=1}$.  The nearest and next-nearest constraint, and the Exclusion Principle constraint, are understood.
Periodic boundary conditions in the space and time dimensions are imposed, and we are able to vary the inverse temperature in lattice units by varying the extension in the time direction.  For the most part we use $N_t=100$, and equal numbers of spin up and spin down particles.  The density is fixed by $n_p$, and for an $L\times L$ lattice $n_p = L^2$ corresponds to half-filling, as in a true fermionic system.

    For numerical simulation of the model via importance sampling, employing the usual Metropolis algorithm, we calculate the change in this action  
  \beq
   \D S = \k\D j + \D V
  \label{dS}
  \eeq 
resulting from a trial update in the trajectory of one of the particles in the system.  We will be interested
in studying the behavior of the system at high density (i.e.\ half-filling and above) as the $U$ and $\k$ parameters are varied.
 
   It should be obvious that despite some similarities, the model of eq.\ \rf{action} is not the Hubbard model. Nor is it a bosonic
field theory.  It is really a hopping model describing a dense set of two types of distinguishable particles 
(``spin up'' and ``spin down'') , with the constraint that no more than one particle of 
either type can occupy any given site of the lattice. There is at least one historical precedent for
treating fermions in that way, namely the original Hartree formulation of the Hartree-Fock approximation in atomic physics.
The Hartree-Fock approximation, of course, consists of writing down the ground state of a set of electrons
moving in a central potential determined by self-consistency.  In the original Hartree version,
the minimal energy state of $N$ electrons is chosen from many body states of the form
\beq
      \Psi(1,2,...,N) = \psi_{s_1}(1) \psi_{s_2}(2)....\psi_{s_N}(N)
\eeq
where $\psi_s(j)$ indicates that particle $j$ is in the energy eigenstate $\psi_s$ of the one-body Schrodinger equation
for an electron moving in a central potential.  The restriction was the Pauli Exclusion Principle:  no two particles could be
in quantum states labeled by the same set of integers $s$.  Of course it was soon noted that this application of 
the Exclusion Principle is insufficient, and that a many-body wave function of this kind is appropriate to distinguishable particles 
rather than identical fermions. The remedy was to impose antisymmetrization via the Slater determinant, and 
this improved method is known as ``Hartree-Fock.''  However, it is worth noting that the original Hartree approximation was
not so terrible at the quantitative level.  Agreement with experiment was certainly improved with 
the appropriate antisymmetrization, but the Hartree version already gives reasonable results for atomic structure, differing 
from the more accurate Hartree-Fock method, in estimates of atomic energy levels, at the 10-20\% level 
\cite{gasiorowicz2007quantum}.\footnote{There are of course instances where the quantitative discrepancy between the
Hartree and Hartree-Fock methods is more severe, e.g.\ when the exchange interaction is crucial, cf.\ \cite{Pfann}.}
The hopping model I have just introduced imposes an exclusion principle on double occupancy, rather than 
on energy eigenstates.  But one might hope that if the particles are reasonably localized 
at the quantum level, then there might still be some resemblance to 
the physics of the Hubbard model, despite the clear violation of Fermi-Dirac statistics.  In any case, given
the absence of any robust computational solution of the Hubbard model away from half-filling, we believe that
the investigation of this simplified (and, as regards identical particle  statistics, evidently wrong) version of that model may still be worth pursuing.   
Of particular interest would be the occurrence of quantum phase transitions in this Euclidean-time quantized hopping model.

\subsection{The transfer matrix}

The connection between a statistical mechanical system in $D$ Euclidean dimensions to a quantum system evolving in real time, in $D-1$ spatial dimensions with unitary time evolution, depends on the existence of a transfer matrix \cite{Creutz:1984mg,Rothe:1992nt}, and the argument goes as follows:  Let the Hilbert space be spanned by a basis $\{|\a\rangle\}$, and we suppose that Euclidean time runs in discrete steps of duration $\e$ from $t=0$ to $t=N_t \e$, with periodic boundary conditions in time.   The statistical mechanics system can be identified with some corresponding quantum mechanical system if there exists a positive Hermitian operator $\T$, known as the transfer matrix, such that
\bea
            Z &=& \sum_\phi e^{-S[\phi]} \non \\
               &=& \tr \T^{N_t}  = \sum_\a \langle \a| \T^{N_t} | \a \rangle
\label{Z}
\eea
where the sum in the first line is over configurations $\phi$, whatever they may be in a given theory. 
If there is such an operator, then the Hamiltonian operator is given by the logarithm, i.e
\beq
           \T = e^{-\e H} ~~~,~~~ H = -{1\over \e} \log(\T)
\eeq
and $H$ must be Hermitian if $\T$ is positive Hermitian, i.e.\ if its eigenvalues are real and positive.  Given the existence of this operator, we define the unitary operator
\beq
           U_\e = e^{-i\e H}
\label{Ue}
\eeq
and so quantum mechanical evolution  of a wave function over $n$ discrete time steps is given by
\beq
           |\psi\rangle_{t_0+n\e} = (U_\e)^n  |\psi\rangle_{t_0} =  e^{-i n \e H} |\psi\rangle_{t_0}
\eeq
One also sees that
\beq
           Z = \tr e^{-N_t \e  H}
\eeq
which gives us the usual connection between the time extension of the periodic lattice, and temperature $T = 1/(N_t \e)$.
Examples of the Euclidean path integral representation of quantum systems at finite temperature include
the 2D Ising model representation of a quantum Ising spin chain \cite{Fradkin:1978th}, and the finite temperature phase transition in a quantized SU(3) gauge theory as detected from the simulation of a Euclidean time path integral \cite{Gattringer:2010zz}.
 Depending on the action, it may or may not be easy to determine the form of $H$ explicitly in the continuum $\e \ra 0$
time limit.  For a single particle moving an external potential one can derive, by the procedure discussed in \cite{Creutz:1984mg,Rothe:1992nt},  the usual $H=p^2/2m + V$
form, with canonical commutation relations $[x,p]=i\hbar$.  For a non-abelian lattice gauge theory, the kinetic term is
a little more complicated, and boils down to a Casimir operator (or, for an SO(3) gauge theory, the familiar squared angular
momentum operator).  For a spin system, it is something else (see e.g.\ \cite{Itzykson:1989sx}).  There is no guarantee that a simple Euclidean
action will lead to an equally simple Hamiltonian in the time-continuum limit, or even that the time-continuum limit exists, 
but what {\it is} guaranteed is that a Hermitian Hamiltonian operator exists if a transfer matrix exists.

    The relevant question is  then whether the hopping model described above has a transfer matrix.  The answer is a qualified yes, and the reason, as in more conventional models, is that couplings in the kinetic term are nearest-neighbor in time.  For the basis in Hilbert space, we can choose eigenstates of particle position $| \{ \bx_n \} \rangle$, where $\bx_n$ is the lattice site occupied by the $n$-th particle in the system, and $s_n = \pm 1$ is the up or down spin of particle $n$, which is fixed from the beginning (e.g.\ particles 1 through $n_p/2$ can be spin up, with the remainder down).  The transfer matrix is then
\bea
\T_{\{\bx'_n\}, \{\bx_m\}}  &=& \langle \{\bx'_n\} |\T|\{\bx_m\} \rangle \non \\
     &=& \exp\left[ - \k \sum_{n=1}^{n_p} J(|\bx'_n-\bx_n|) \right. \non  \\
     &  &  \left.  - \oh \sum_{n \ne m} \left(V_{nm}(\bx_m,\bx_n) + 
                    V_{nm}(\bx'_m,\bx'_n)\right)/2 \right] \non \\
\eea
where
\bea
            J(|\bx' - \bx|) &=&  \left\{ \begin{array}{cl} 
                   0 & \mbox{if~} |\bx'-\bx| = 0 \cr
                   1 & \mbox{if~} |\bx'-\bx| = 1 \mbox{~or~} \sqrt{2} \cr
                \infty& \mbox{otherwise} \end{array} \right.
\label{J}
\eea
and
\bea             
            V_{nm}(\bx,\by) &=&  \d_{\bx,\by} \times \left\{ \begin{array}{cl} 
                               U & \mbox{if~} s_m \ne s_n  \cr  
                            \infty &  \mbox{if~} s_m = s_n      \end{array} \right.
\label{V}
\eea

The matrix $\T$ is clearly both real and symmetric, and satisfies \rf{Z} for the Euclidean action described above,
where the infinities in \rf{J} and \rf{V} simply implement the nearest and next-nearest hopping constraint, and the Exclusion
constraint.  Positivity is more difficult to prove (hence the term ``qualified''), but seems very plausible for the following reasons:  Denote the eigenstates and eigenvalues of the transfer matrix by $\psi_k$ and $\l_k$ respectively.  Then the thermal expectation
value of any operator $Q[\{x_n\}]$ is given by
\beq
            \langle Q \rangle = {1\over Z} \sum_k \langle \psi_k|Q|\psi_k \rangle (\l_k)^{N_t}
\eeq
Suppose $Q$ is positive for any value of its arguments.  If some of the $\l_k$ are negative, then the thermal expectation value
is not necessarily positive, for all possible operators of this type, at odd values of $N_t$. But we can see from the Boltzmann distribution in \rf{Z} that {\it any} positive $Q$ must have a positive thermal expectation value,  for any choice of $\k,U$ and any $N_t$. This positivity property is not limited to equal time operators, but holds for any positive function of the degrees of freedom at all times (i.e.\ the particle trajectories) $Q[\{x_n(t)\}]$, again for any choice of $\k,U,N_t$.  For this reason the positivity of $\T$ itself seems very plausible (the author does not know of any counterexamples), and 
will be assumed here. 

    Note that a Hermitian Hamiltonian and unitary state evolution follows from the existence of the transfer matrix; the continuous time limit is not required.  For a particle moving in a potential in three continuous space dimensions, or a lattice gauge theory on a three dimensional lattice, the transfer matrix formulation leads to real time evolution in discrete time steps. In both cases it is possible (although in the second case not completely straightforward), to take the continuous time limit, and arrive at a Hamiltonian containing differential operators.  It would be interesting to study whether a continuous time limit, for the simple model I have proposed, could also be achieved by restoring dimensions to the lattice couplings, and scaling them in such a way that the physics is preserved in the continuous time limit.  But this is beyond the scope of my present article.

\subsubsection{Ergodicity}

  We numerically simulate the system we have described via the standard Metropolis algorithm. 
This algorithm is an example of a Markov chain simulation of a quantum system at finite temperature in which, beginning from some initial configuration in Euclidean spacetime, the system evolves along the Markov chain towards (if all goes well)
a Boltzmann probability distribution of the Euclidean action.  For a system with a transfer matrix $\T$ this is also a simulation
of the statistical mechanics of a quantum mechanical system at finite temperature.  The Metropolis algorithm,  together with variants of the method such Heat Bath and Hybrid Monte Carlo, are collectively known as ``importance sampling.''  Another way of generating an appropriate sequence of configurations in fictitious time, currently under investigation for possible application to the sign problem \cite{Berger:2019odf}, is the Langevin equation \cite{Damgaard:1987rr}.  Importance sampling is the numerical technique which underlies, e.g., the successes obtained in lattice QCD by the high energy physics community.  It is also the method used to investigate the hopping model I have described above.  For a discussion of Markov processes and importance sampling, mainly in the context of non-abelian lattice gauge theory, cf.\ Gattringer and Lang \cite{Gattringer:2010zz}. Of course an assumption which underlies the method of importance sampling is the ergodicity of the system itself.

     We may think of the Euclidean path integral as representing the dynamics of a discretized system of trajectories (which might be thought of as ``fibers'' of some kind) with repulsive interactions and the Exclusion Principle just mentioned, in contact with a heat bath.   Given the existence of a transfer matrix, the Euclidean path integral also represents the canonical ensemble for the dynamics of a quantum system at finite temperature. But following the line of thought that the Euclidean path integral represents the statistical mechanics of a system of interacting fibers, we may consider the following possibility:  when the fibers are dense and repel one another, the system might have the characteristics of a glassy polymer of some kind, with a corresponding breakdown of ordinary ergodic behavior \cite{mauro}.  If so, this would be a characteristic of the system itself, rather than the choice of algorithm used to simulate the system.   Since the Euclidean theory corresponds, according to the usual arguments, to a quantum theory at finite temperature, and supposing that non-ergodic behavior is encountered in the statistical system of particle trajectories in Euclidean time, a natural question is how this behavior would be manifested in the real time quantum mechanics of the system.  The most natural manifestation, we believe, would be the breakdown of the eigenstate thermalization hypothesis.   

Suppose, at the classical level, that the time averages of a system in contact with a heat bath do not reproduce the canonical ensemble, and this will be the case if the time evolution is non-ergodic. We may then be fairly confident that removing the stochastic influence of a thermal bath, and allowing the system to evolve in isolation, will not restore ergodicity.\footnote{One can imagine very special cases where the isolated system is integrable and non-ergodic, but becomes ergodic when placed in contact with a heat bath.  But the opposite situation is unlikely.  It is hard to see how, if the system in contact with a heat bath is non-ergodic, the isolated system, undisturbed by random thermal influences, could nonetheless contrive to be ergodic.}  At the quantum level, eigenstate thermalization ensures that the real time evolution of an isolated quantum system will reproduce the canonical ensemble represented by the Euclidean path integral.  But if the statistical system described by that path integral is glassy and non-ergodic, and the system inevitably gets ``stuck'' in some region of phase space which depends on the initial conditions, then the canonical ensemble cannot represent the evolution of this system (as described by some Markov process) in contact with a heat bath.  Supposing, then, that the evolution of a quantum system in contact with a heat bath is non-ergodic, one would not expect ergodicity to be restored when the randomizing influences of a heat bath are removed, and the isolated system simply evolves in real time under Schrodinger evolution. This strongly suggests a breakdown of eigenstate thermalization.
  
\subsection{\label{sec:leve2} Observables of the hopping model} 

In the present case, the importance sampling procedure is to go time slice by time slice, at each time updating the location of each particle on the two-dimensional lattice. In the Metropolis update for each particle we choose at random a trial hopping direction (to a nearest or next-nearest neighbor site), and measure the change in the number of hops $\D j$ along the trajectory of the given particle (this must be in the range $-2 \le \D j \le 2$), and the change in total potential energy of the system as the particle is moved from one site to another.  This gives $\D S$ in \rf{dS}, and the change is then accepted or rejected in the usual way. The trial move is constrained by the Exclusion Principle, and the restriction that a change in particle position at time $t$ should not result in a hop from time $t-1$ to $t$, or time $t$ to $t+1$,  such that the distance between sites at the earlier and later times exceeds the nearest or next-nearest neighbor hopping limit.
 
  The two dimensional lattice size of the simulations has been varied from  $10\times 10$ to $50\times  50$ sites 
  in order to study the volume dependence of the observables. Because there are
  equal numbers of each type of particle, the maximum number of particles allowed on an $L\times L$ lattice is
  $2 L^2$. 
    
   In our computation, we calculated (i) the hop susceptibility;  and (ii) the probability that a 
   particle remains in the initial site after a Euclidean time lapse $t$.  These are computed as a function of $\k$ 
   for various $U$ values, inverse temperature (the extension of the lattice in the inverse time direction), and for different densities.
   
   \bigskip
   
   Let us define the average hopping number at time $t$
   \beq
         \text{hop}(t) = {1\over n_p} \sum_{n=1}^{n_p} \oh( j(n,t-1) + j(n,t))
         \label{hop}
   \eeq
   and the corresponding hopping susceptibility
   \beq
      \chi_{hop} =  n_p {1\over N_t} \sum_{t=1}^{N_t} (\langle \text{hop}(t)^2\rangle - \langle \text{hop}(t)\rangle^2)
      \label{jsus}
   \eeq   
 This observable has been defined so as to be local in time, and in a real time quantization could
 presumably be computed, at zero temperature, from the ground state wave functional.
    
    In order to investigate localization, we define
    \beq
    \begin{split}
      n(t_0, t_0 +t) = ~ & \text{no. of particles which are at the same} \\
                                & \text{lattice site at times $t_0$ and $t_0 + t$} 
     \end{split}
     \eeq
   and from this quantity we compute the probability that a particle will remain in the same position for $t$ time steps
   \beq
   P(t)  = \left\langle  \frac{1}{N_t}\sum_{t_0 = 1}^{N_t}\frac{n(t_0,t_0+t)}{n_p} \right\rangle
   \label{frac}
   \eeq  
   
 \section{Results}

\subsection{Non-ergodicity at half-filling}

  For half-filling (particle density 50\% of maximum), we computed observables with two different starting configurations, which we term  ``random'' and ``minimum energy'' (or just ``minimum'') respectively.  Initially all particle trajectories are constant in time, but in the random configuration the $x,y$ position of each trajectory is chosen at random, apart from the constraint of the Exclusion Principle.  So with this initialization a certain fraction of sites are doubly occupied at each time.  In the minimum energy configuration there is one particle per site, and no doubly occupied sites, with up and down spins alternating in an antiferromagnetic pattern.  In these initial configurations we have $K(n)=0$ for all particles, and the potential energy vanishes in the minimum configuration.
    
\begin{figure}[htbp] 
   \centering
    \includegraphics[width=3in]{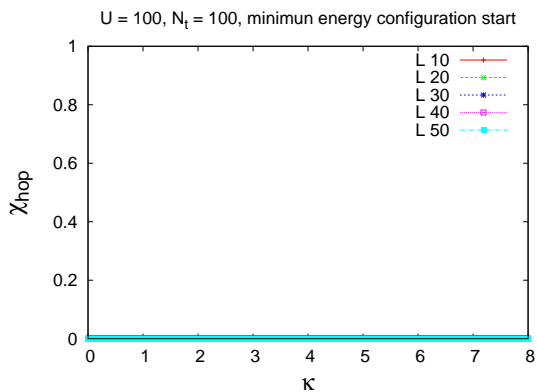} 
    \caption{A trivial result: hop susceptibility ($\chi_{hop}$ of eq.\ \rf{jsus}) vs.\ $\k$  at low temperature ($N_t$ = 100),
      strong repulsion $U = 100$, and various spatial areas $L^2$, for 50\% density, initialized at the minimum energy      configuration.}
    \label{fig3}
 \end{figure}

  Let us first show the hop susceptibility result at $U = 100$ with the minimum and random initializations. With an initial
  minimum energy configuration there is effectively no hopping at all, due to the strong Coulomb repulsion.  The hop susceptibility $ \chi_{hop}$ must obviously be zero for all $\k$ values, at such a large value of $U$, and of course that is what one  finds (Fig.\ \ref{fig3}).
 
     \begin{figure}[htbp] 
   \centering
    \includegraphics[width=3in]{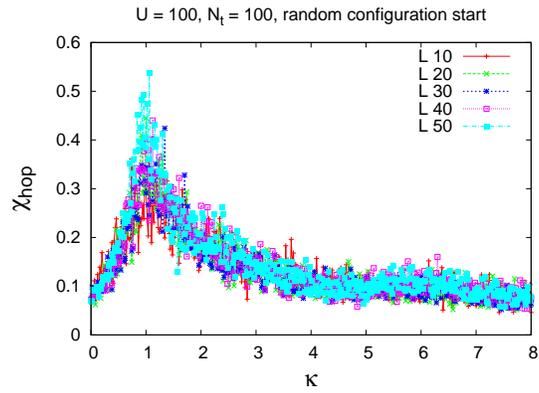} 
    \caption{$\chi_{hop}$ vs.\ $\k$ at low temperature ($N_t$ = 100), strong repulsion ($U = 100$),
     at various spatial areas $L^2$, 50\% density with a random initial configuration.}
    \label{fig1}
 \end{figure}
 
  \begin{figure}[htbp] 
   \centering
    \includegraphics[width=3in]{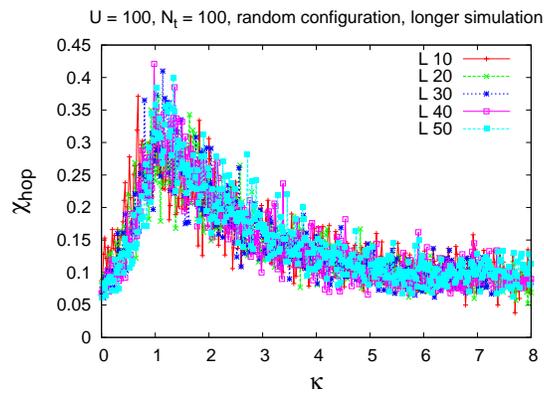} 
    \caption{Same as Fig.\ \ref{fig1} but with both thermalizing and subsequent Monte Carlo sweeps increased by a factor of 10, to at total of $10^5$ sweeps.  This seems to make little difference to the final results.}
    \label{fig2}
 \end{figure}

  \begin{figure}[htbp] 
   \centering
   \subfigure[~]{
    \includegraphics[width=3in]{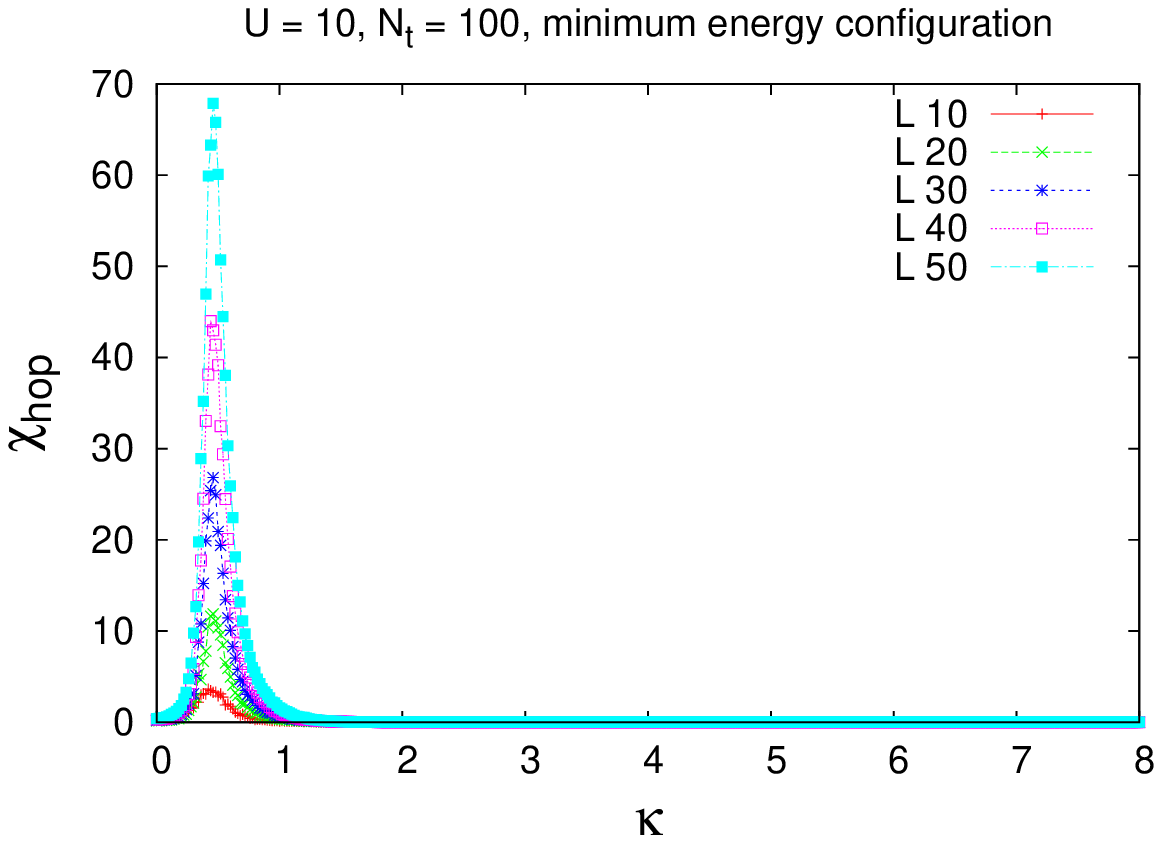} 
    \label{fig18}
   }
  \subfigure[~]{
    \includegraphics[width=3in]{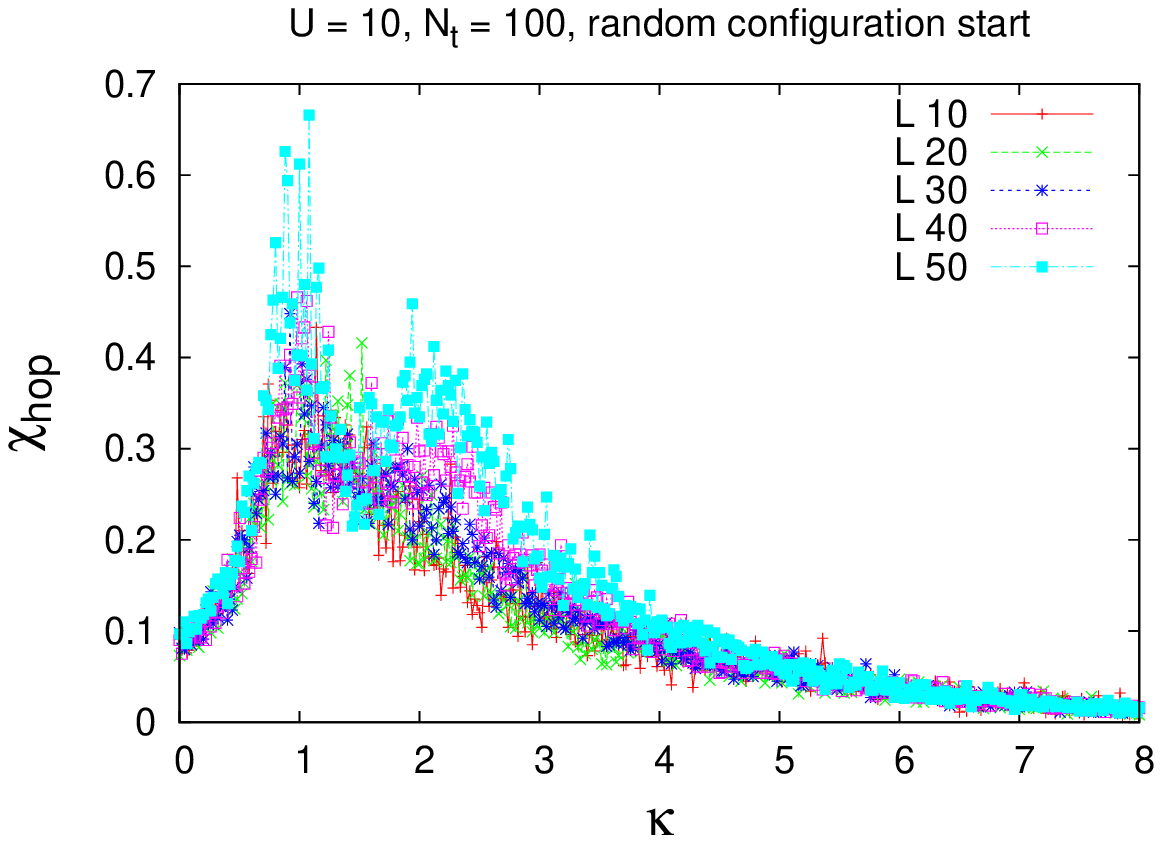} 
    \label{fig17}
   }
   \subfigure[~]{
    \includegraphics[width=3in]{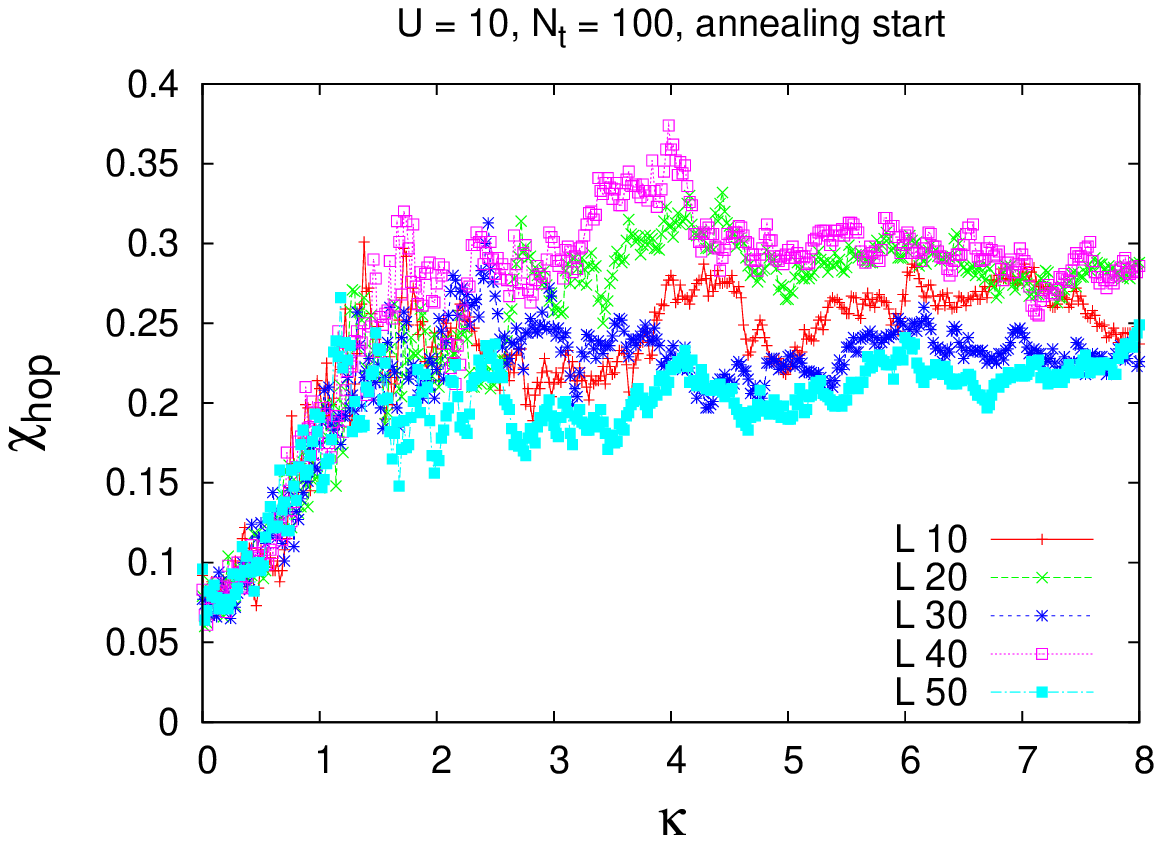} 
    \label{fig19}
    }
  \label{weaker50}
  \caption {$\chi_{hop}$ vs $\k$ at weaker repulsion $U=10, ~N_t = 100$, 50\% density at spatial areas $L^2$.
    Initialization in subfigures at: (a) minimum energy; (b) random; (c) annealing start configurations.}
 \end{figure}
  
     On the other hand, in the random configuration start, the situation is dramatically different. Data was taken on $L\times L$
  lattices for $L =10$ up to $L=50$, and $N_t=100$. In this situation there is certainly some amount of hopping, but 
  the particles fail to find their way to the minimum energy configuration.  Figure \ref{fig1} shows the results for  
  $\chi_{hop}$ at $U=100$ obtained in a numerical simulation with
 1000 thermalizing sweeps, followed by 9000 sweeps, with data taken every 100 sweeps.  One might wonder if the system
 simply needs more Monte Carlo time to find its way to the minimum energy configuration.  In Figure \ref{fig2} the number
 of thermalizing and data taking sweeps have been increased by an order of magnitude, with very little change in the hop
 susceptibility.  Presumably, at a sufficiently large number of sweeps on a finite lattice, eventually the minimal energy configuration would be obtained, if there is any allowable path to that configuration.
 
   We think it is likely however, given these results, that the number of sweeps required to reach a minimum energy configuration would rapidly diverge
 with volume, while keeping the density at half-filling fixed.
  
     At a lower value of $U=10$ there can be particle hopping even with a minimum energy initialization,  as seen in Fig.\ \ref{fig18}.
Since the height of the susceptibility peak is volume dependent, there is an apparent suggestion here of a quantum phase
transition.   It is not hard to characterize the phase on the right side of the peak in Fig.\ 4(a), i.e.\ the high-$\k$ region.  This is a ``minimal energy'' phase, in which the particle positions are more-or-less fixed at the initial minimum energy starting configuration, as in Figure 1, and fluctuations away from this initial configuration are very strongly suppressed.  That is consistent with the value of $\chi_{hop}$ to the right of the peak, which rapidly drops with increasing $\k$ below $10^{-3}$ at $L=50$.  In contrast, the susceptibility to the left of the peak is non-zero down to $\k=0$, where $\chi_{hop} \approx 0.4$ at $L=50$, although this non-zero limit at $\k=0$ is not obvious because of the scale of the $y$-axis of the plot.  Thus to the left of the peak, particle trajectories can fluctuate in the course of the simulation, and in the process it is possible to deviate substantially from the initial minimum energy configuration.  There seems to be a very sharp transition between the fluctuating and minimal energy phases at the location of the peak in Fig.\ 4(a), providing the system is always initialized to the minimum energy trajectories. Matters are different at high $\k$, with a random initialization (Fig.\ 4(b)), where there is clearly no transition to a minimal energy phase.  The reason must be that with a random start the system is unable to find its way to the minimum energy configuration (where it would be stuck), and as a result there is no sharp transition between the low and high $\k$ regions.  The dependence of $\chi_{hop}$ at large $\k$ on the initialization is, of course, evidence of non-ergodic behavior.
 
  In addition to the random and minimum initial configurations, we have also investigated what could be described as  ``annealing''
initializations.  For both random and minimum  configurations, we initialize to a
random or minimum configuration at each $\k$ value, before evolving the system via the Metropolis algorithm.
Instead, for the annealing initialization, the system is only initialized, with a random initial configuration, at $\k=0$.  After computing the hop susceptibility at $\k = 0$, we use the configuration obtained at the last Monte Carlo sweep as the initial configuration at
 the next value of $\k$, i.e.\ $\k = \d \k$, with $\d \k = 0.02$.  The last configuration of the simulation at that $\k$ value is then used as the initial
configuration at the next $\k$ value at $\k = 2 ~\d\k$, and so on, with the last configuration at $\k = n ~\d\k$ used as the initial configuration for the
$\k = (n+1) ~\d\k$ simulation, over the full range of $\k$ used in the calculation.

The data with annealing initialization at $U = 10$ and 50\% density produced the surprising results shown in Fig.\ \ref{fig19}. The reason this is surprising is that we see a substantial hopping susceptibility even at rather large $\k$, which should have suppressed all hopping.  But this is clearly an effect of the initialization.  At small values of $\k$ it is not surprising that there will be some non-negligible number of jumps along any given trajectory.  But since the last configuration at $\k = n \d\k$ is the initial
configuration of the next simulation at $\k = (n+1) \d\k$, there is the possibility that the multiple jumps found in trajectories at
low $\k$ become ``frozen in'' as $\k$ is raised to higher values, and this appears to be what happens.  At large $\k$ the energetics would prefer a low $K(n)$ value, with the minimum $K(n)=0$ value obtained for a trajectory with no jumps whatever.  The persistence of a hopping susceptibility at large $\k$ simply indicates that the system of trajectories cannot find its way, at large $\k$, to anywhere near that minimum, and at large $\k$ there may be fluctuations, in a rough trajectory, in which an increase in kinetic
energy is compensated for by a decrease in potential energy, or vice versa; there may even be fluctuations which do not alter
either energy.   We will further discuss the ``freezing in'' of multiple jumps in section \ref{peak2}.
 
 \begin{figure}[htbp] 
  \centering
   \subfigure[~]{
    \includegraphics[width=3in]{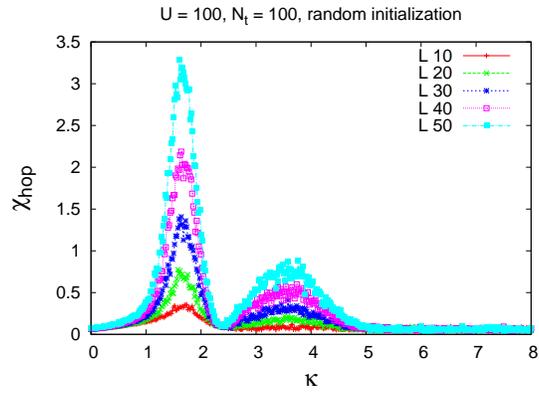}  
    \label{fig4}
    }
    \subfigure[~]{
    \includegraphics[width=3in]{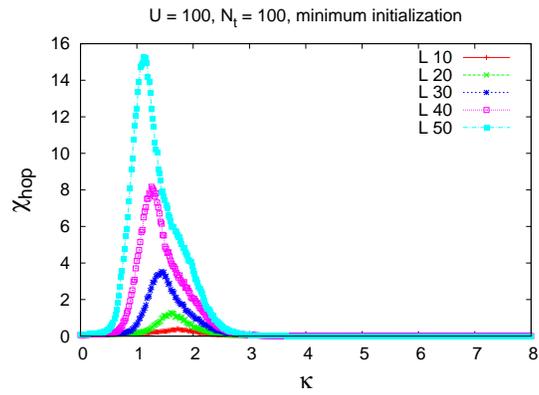} 
    \label{fig5}
    }
    \centering
    \subfigure[~]{
    \includegraphics[width=3in]{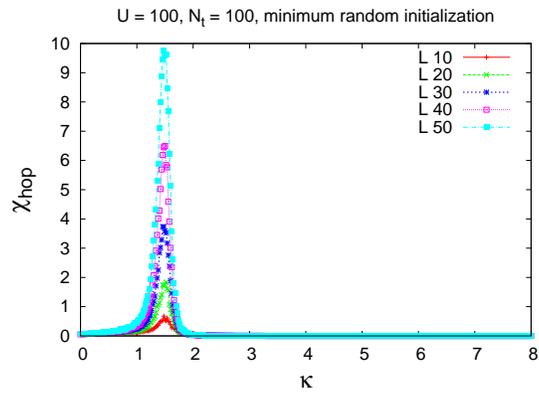} 
    \label{fig6}
    }
    \caption{$\chi_{hop}$ vs $\k$ at 60\% filling.  Strong repulsion $U=100$, low temperature $N_t=100$, spatial areas $L^2$.  Initialization in the subfigures is at (a) random; (b) mimimum; (c) minimum random
    configurations.}
    \label{fig456}
 \end{figure}

  These results at both $U=100$ and $U=10$, which are so very clearly dependent on the starting configuration, are a first indication of non-ergodicity (or exceptionally long relaxation times) in the classical system of particle trajectories, and a corresponding non-ergodicity in the associated quantum system of point particles.

\subsection{Non-ergodicity at 60\% filling}
\subsubsection{U = 100}

  We now increase the particle density to 60$\%$ of maximum, and compute the hop susceptibilities 
for random, minimum, and annealing initializations.
At 60\% we have investigated two types of minimum energy initializations.
After placing the up and down spins alternately on each site, so that there is one particle at each
lattice site, we can place the remaining particles at random (with the constraint of the Exclusion Principle),
to form doubly occupied sites distributed randomly in the lattice.  We will refer to such an initialization as
``minimum random''.  An alternative is to add remaining particles in one corner of the lattice, so that the
doubly occupied sites are located in one connected region of the lattice.  We will refer to this initialization as
simply ``minimum.''

\begin{figure}[htbp] 
   \centering
    \includegraphics[width=3in]{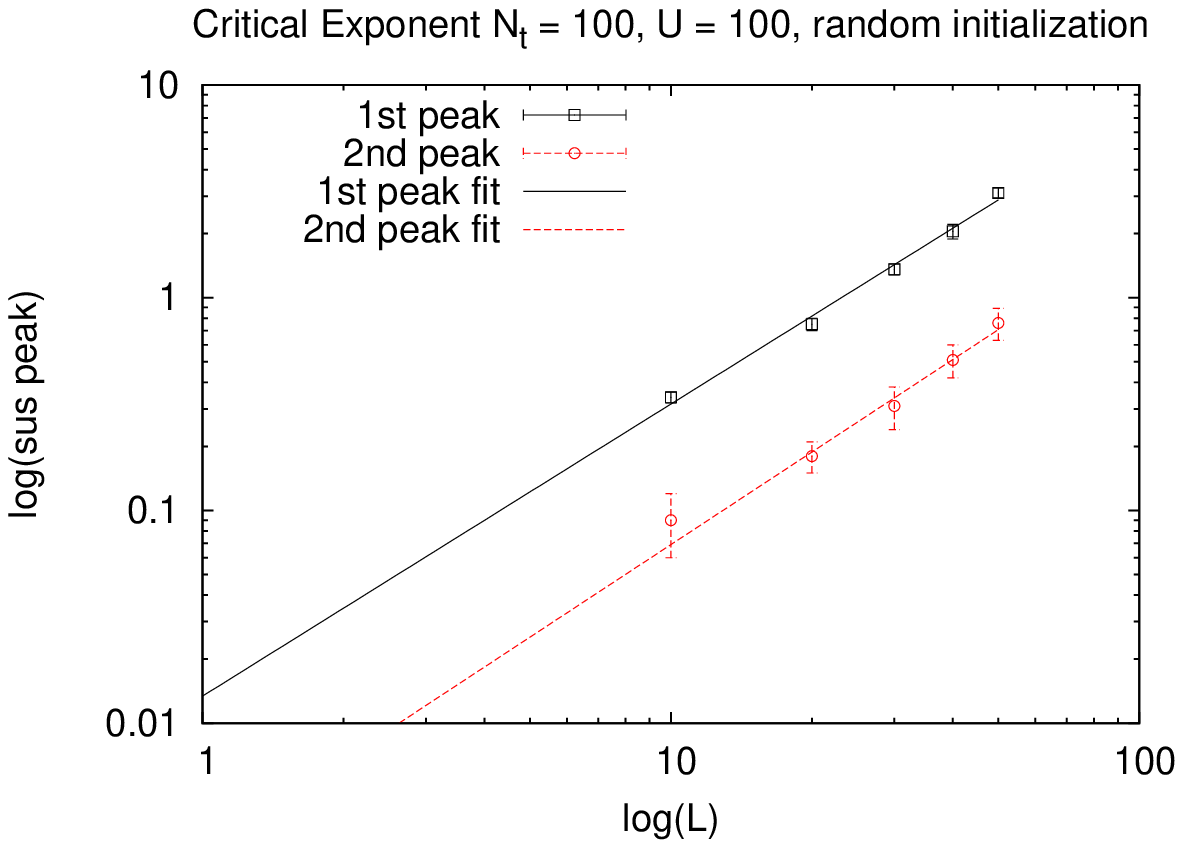}
    \caption{critical exponent fittings for low temperature (T = 100) hopping susceptibility at 
    U = 100, 60$\%$ filling random initial configuration}
    \label{fig7}
 \end{figure}

 \begin{figure}[htbp]
\subfigure[~]  
{   
 \label{8a}
 \includegraphics[width=2.5in]{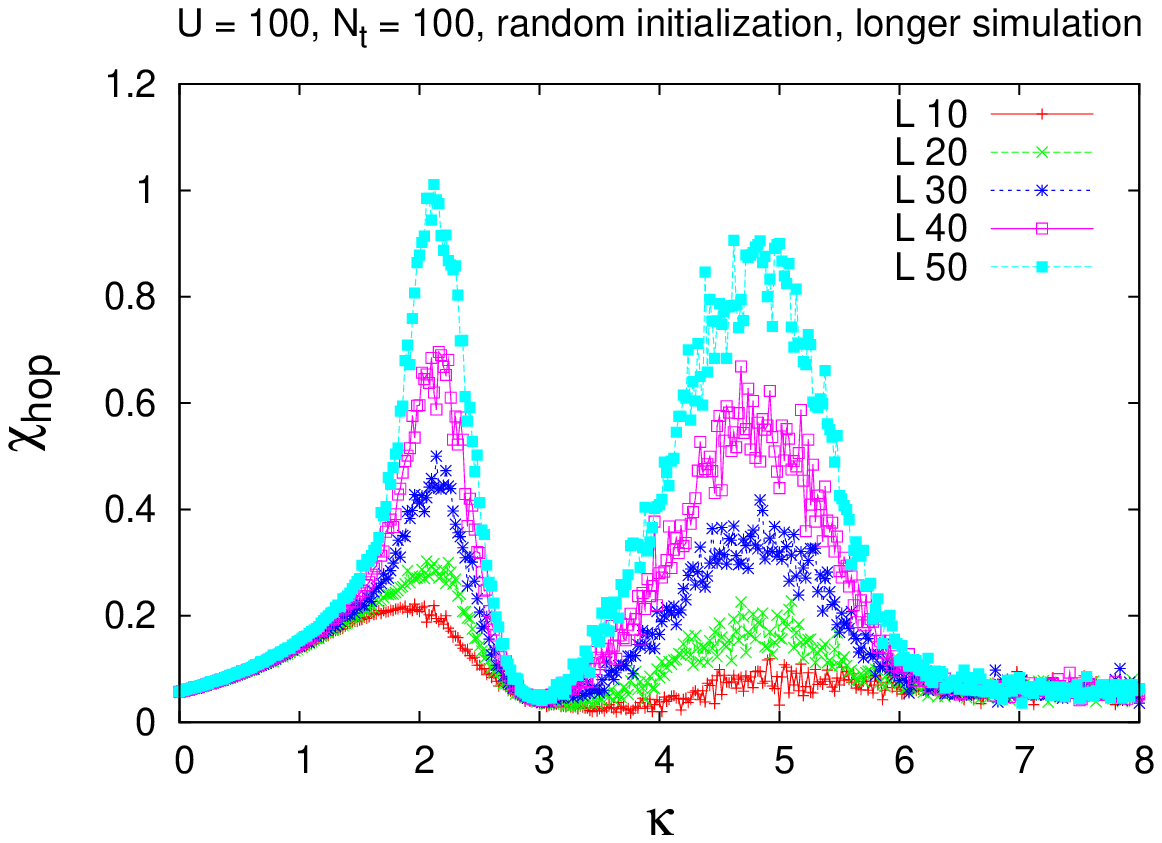}
}
\subfigure[~]  
{   
 \label{8b}
 \includegraphics[width=2.5in]{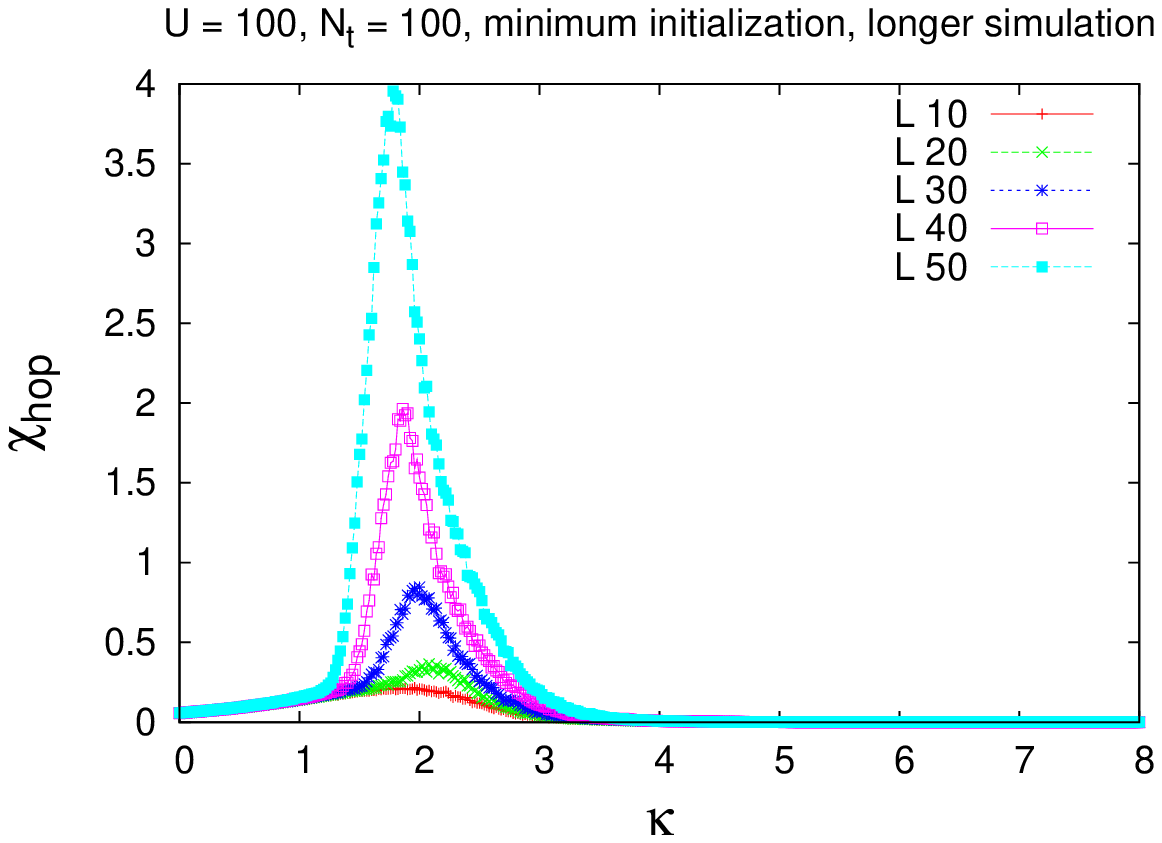}
}
\centering
\subfigure[~]  
{   
 \label{8c}
 \includegraphics[width=2.5in]{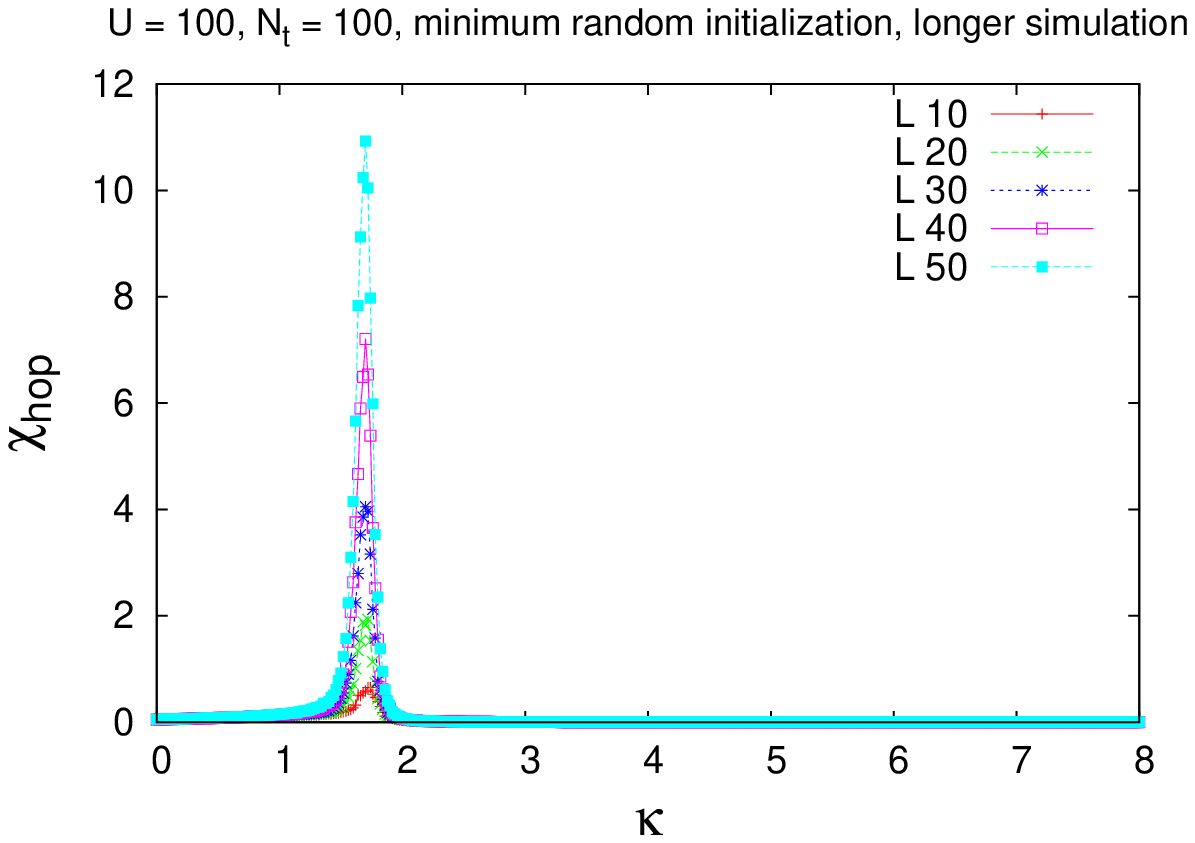}
}
\caption{Same as Figure \ref{fig456}, but with both thermalizing and subsequent Monte Carlo sweeps increased by a factor of 10, to at total of $10^5$ sweeps. Initialization at (a) random (b) minimum; (c) minimum random starting configurations.}
\label{fig8}
\end{figure}

It is interesting that at 60\% filling, with random, minimum, and minimum random initializations, we seem to
see (judging from the hop susceptibility) evidence of a quantum phase transition.  Yet the plots of susceptibilities
differ for the three initializations.  In Fig.\ \ref{fig4}, the random initial particle configuration shows there are two peaks which both grow with spatial volume, indicating two quantum phase transitions,  but with both minimum initialization Fig.\ \ref{fig5}) and minimum random initialization (Fig.\ \ref{fig6}) there is only a single peak.  The second peak in  Fig.\ \ref{fig4}, which appears only for random initialization, and hopping probabilities beyond the second peak, will be discussed in section \ref{peak2} below.
  
 We have computed the  critical exponent from finite size scaling for the two peaks seen in Fig.\ \ref{fig4}.  The maximum peak height vs.\ volume is displayed on a log-log plot in Fig. \ref{fig7}, and a best fit of $(\chi_{hop})_{max}$ to $a L^{(\gamma/\nu)}$, where $a$ is a constant, gives $(\gamma/\nu)_{first}$ = 1.37$\pm$ 0.07, $a$ = $0.013\pm0.003$ for the first peak and $(\gamma/\nu)_{second}$= 1.45$\pm$0.11, $a$ = $0.003\pm0.001$ for the second peak.
  
As before, we check if our calculation depends on the number of Monte Carlo sweeps.  The corresponding results obtained by increasing both thermalization and subsequent update sweeps by a factor of 10 are shown in Fig. \ref{fig8}.  The peaks remain in hop susceptibilities and their positions have shifted slightly.  But while the peak heights for the random and minimum initializations have decreased, the peak height for minimum random is about the same after increasing the number of sweeps by an order of magnitude.

 \begin{figure}[htbp] 
   \centering
    \includegraphics[width=3in]{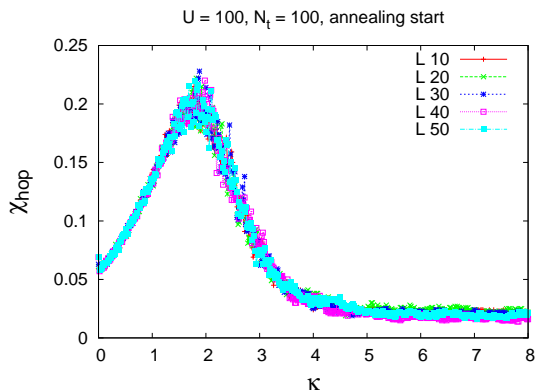} 
    \caption{Low temperature ($N_t$ = 100) hopping susceptibility as a function of $\k$, $U = 100$, 60$\%$ density with the annealing initialization.}
    \label{fig9}
 \end{figure}

\begin{figure}[htbp]
\subfigure[~minimum]  
{   
 \label{11a}
 \includegraphics[scale=0.51]{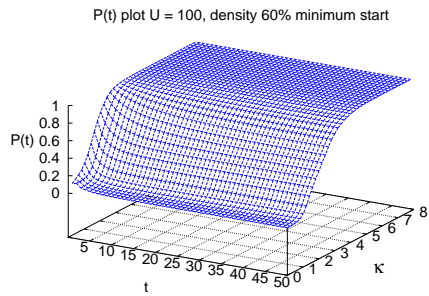}
}
\subfigure[~minimum random]  
{   
 \label{11b}
 \includegraphics[scale=0.51]{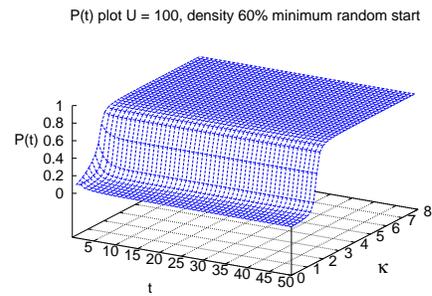}
}
\centering{
\subfigure[~annealing] 
{   
 \label{11c}
 \includegraphics[scale=0.51]{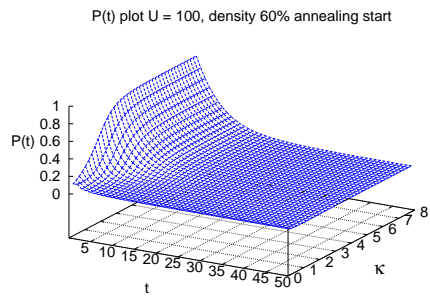}
}}
\caption{Localization probability $P(t)$ vs.\ $t$ and $\k$ on a $50\times 50$ spatial lattice at 60\% filling, at $U=100, N_t=100$ and spatial areas $L^2$.  Initialization at (a) minimum;
(b) minimum random; (c) annealing start configurations.} 
\label{fig11}
\end{figure}

The most dramatic difference is seen with the annealing initializations, with hop susceptibilities displayed in Fig.\ \ref{fig9}.
With this initialization there is essentially no dependence of peak size on lattice area, and no evidence whatever 
of a quantum phase transition.   But we observe that the data on a $10\times 10$ lattice, with random, mimimum, and minimum random initializations, is not so very different from the annealed case.  The suggestion is that non-ergodicity, or, at least, very long relaxation times, is a phenomenon which increases very rapidly with lattice size.  Presumably this translates, on a very large lattice, to a complete breakdown of  eigenstate thermalization in the corresponding real-time quantum system of point particles on the lattice.  In fact the situation is not so different from what we have already seen at half-filling.  Given a sufficiently lengthy simulation, the system must eventually find the minimal energy configuration.  But for a sizable lattice, the required simulation time to reach that minimal energy is probably beyond the reach of any realizable computation.

Our second observable is  the localization probability $P(t)$, which represents the probability that a particle at some time $t_0$,
at a location $(x_0,y_0)$, will be found at that same site after $t$ units of Euclidean time, i.e.\ at time $t_0+t$.  The results,
for the mimimum, minimum random, and annealing initializations are shown in Fig.\ \ref{fig11} for a $50\times 50$ spatial
lattice.  A $P(t)$ which reaches a plateau at large $t$ is indicative of strong localization, and this is what is seen at the larger $\kappa$ values in Figs. \ref{11a} and \ref{11b}.  The sudden onset of localization with increasing $\kappa$ is especially apparent in the minimum random
initialization (Fig.\ \ref{11b}). Note that the positions of the first peak in each plot in the hop susceptibilities in Fig.\ \ref{fig5} and Fig.\ \ref{fig6} approximately coincide with the positions of the onset of the particle localization. In contrast, with the annealing initialization (Fig.\ \ref{11c}) we see $P(t) \ra 0$ with increasing $t$ even at large $\k$, but this is not hard to understand.  In the annealed case (here, and in all calculations of $P(t)$ we use $\d \k = 0.2$), the particle trajectories at larger $\k$ are initialized to disordered configurations that are inherited from the simulations at small $\k$.  Thus there appears to be no localization, even if (as must be the case at large $\k$) particle hopping occurs only rarely in the course of the simulation.

\begin{figure*}[htbp]
\subfigure[~]
{   
 \label{10a}
 \includegraphics[scale=0.55]{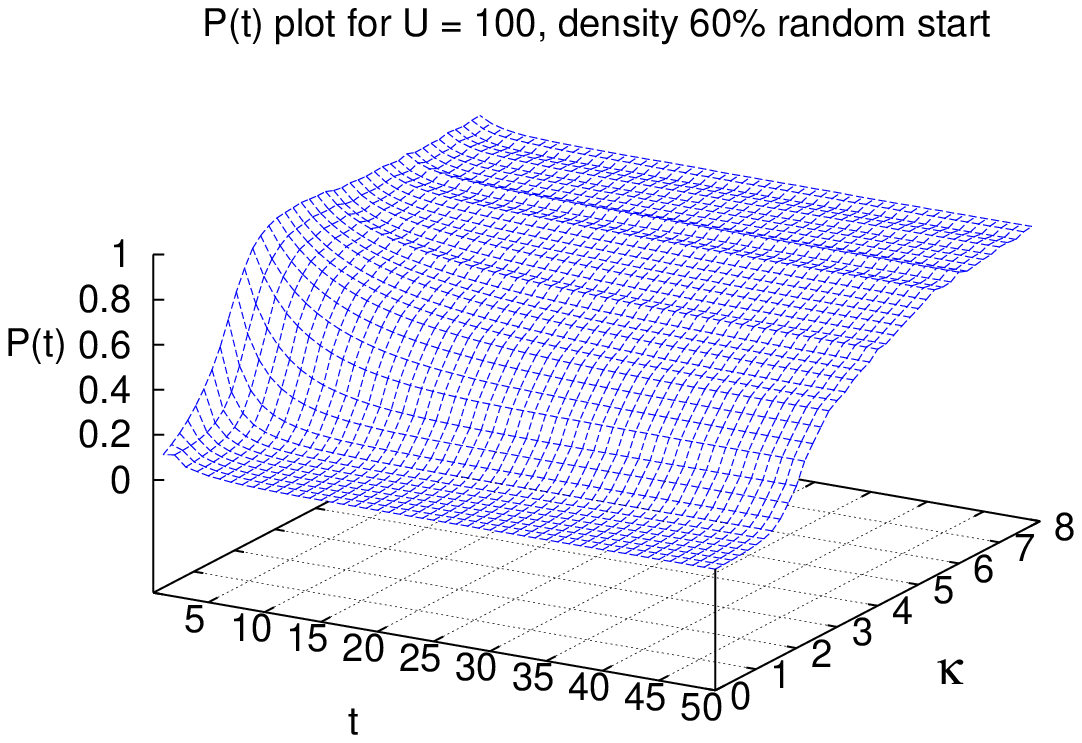}
}
\subfigure[~]  
{   
 \label{10b}
 \includegraphics[scale=0.55]{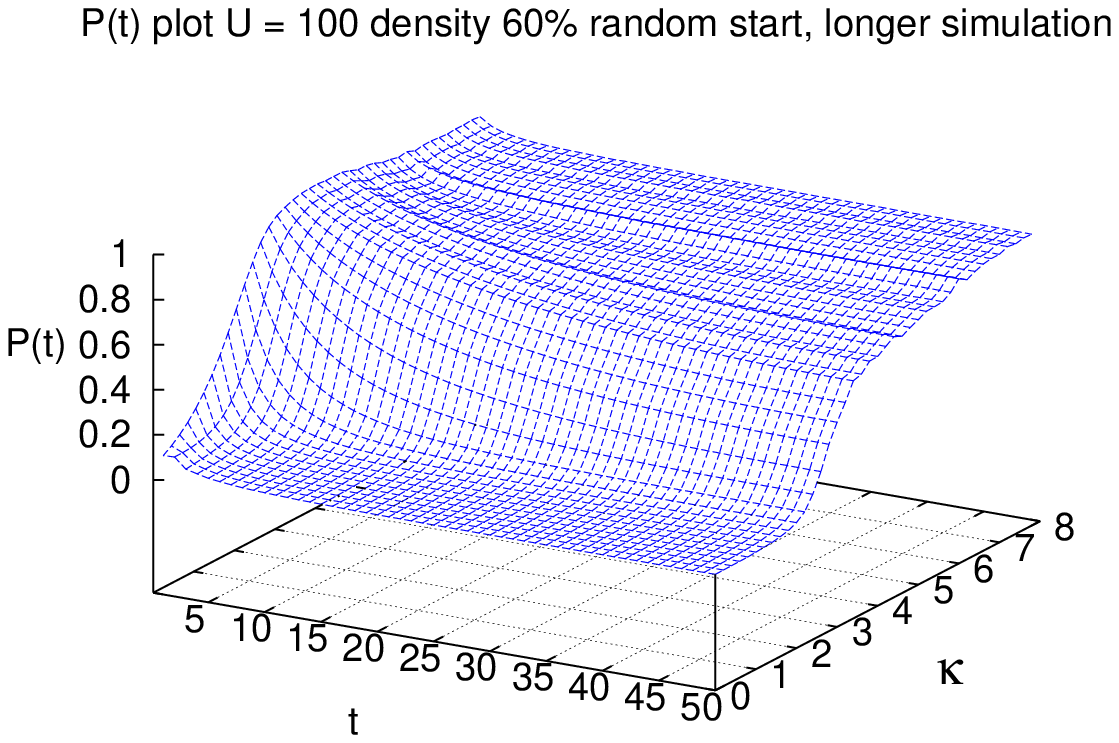}
}
\caption{Same as Fig.\ \ref{fig11} for random initializations.  (a) $10^4$ total Monte Carlo sweeps; (b) $10^5$ total Monte Carlo
sweeps.} 
\label{fig10}
\end{figure*}

The plots shown in Fig.\ \ref{fig10} are for the random initialization, where the onset of localization with $\k$ is
not quite as abrubt.  Fig.\ \ref{10a} shows the numerical results for 1000 thermalizations followed by 9000 sweeps, as usual, with
data taken every 100 sweeps.  Fig.\ \ref{10b} is the same computation, but with thermalization and subsequent Monte Carlo sweeps increased by a factor of 10; this increase seems to make little difference to the result.   Again we find that the onset of localization with $\k$ seems to coincide with the start of the first peak in hop susceptibility.   Overall, it appears that the quantum phase transitions seen in Figs.\ \ref{fig456} and \ref{fig8} (at least the first peak, in the case of random initialization) are associated with a localization transition of some kind, reminscent of a glass transition.

The localization data for the annealing initial configuration is shown in Fig.\ \ref{11c} where we see that $P(t) \ra 0$ after a short time $t$ interval, indicating a lack of localization, and therefore no abrupt transition to localization. This is consistent with our hop susceptibilities calculations in Fig. \ref{fig9} where we found that no phase transition occurred.

\subsubsection{$U = 1, 0$}

   We have repeated the previous computations  for ${U = 10, 1, 0}$ at 60\% density; the results for $U=10$ are qualitatively
quite similar to the previous $U=100$ case.  
  
\begin{figure*}[htbp]
\subfigure[~random]  
{   
 \label{22a}
 \includegraphics[scale=0.5]{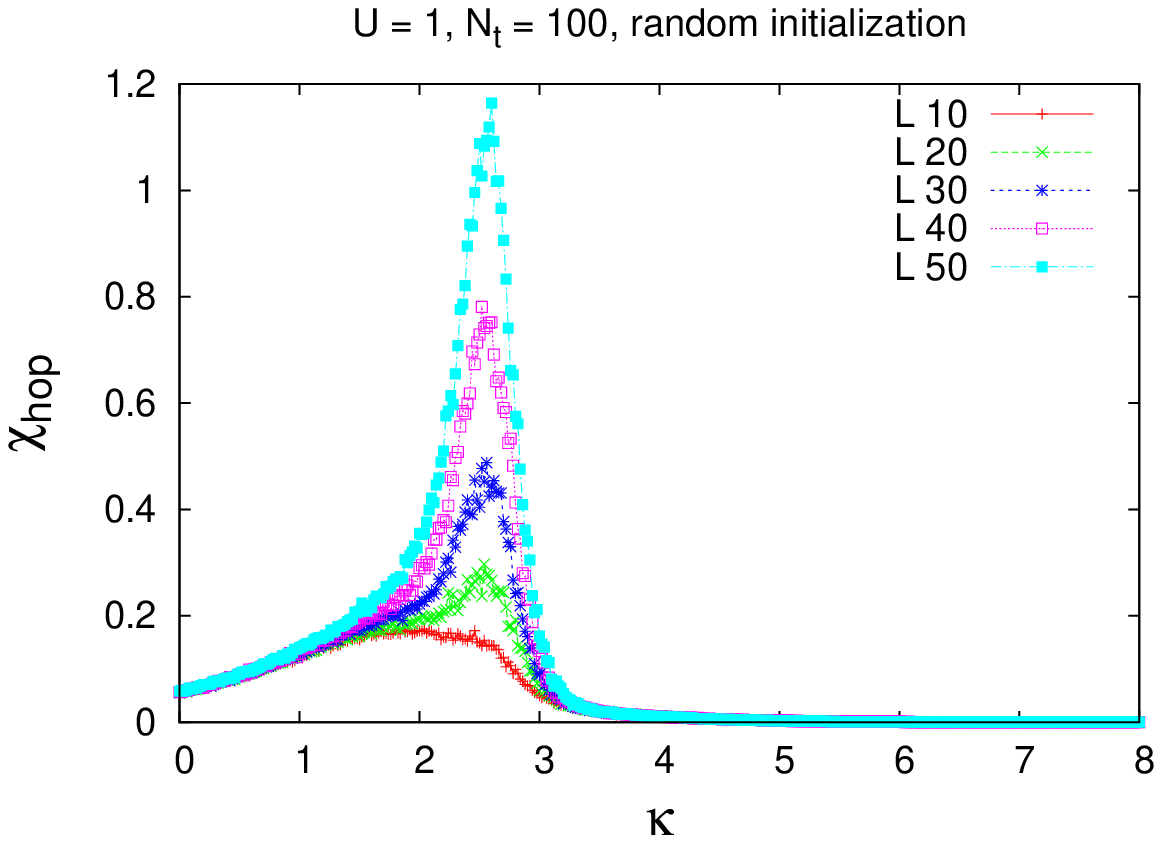}
}
\subfigure[~minimum]  
{   
 \label{22b}
 \includegraphics[scale=0.5]{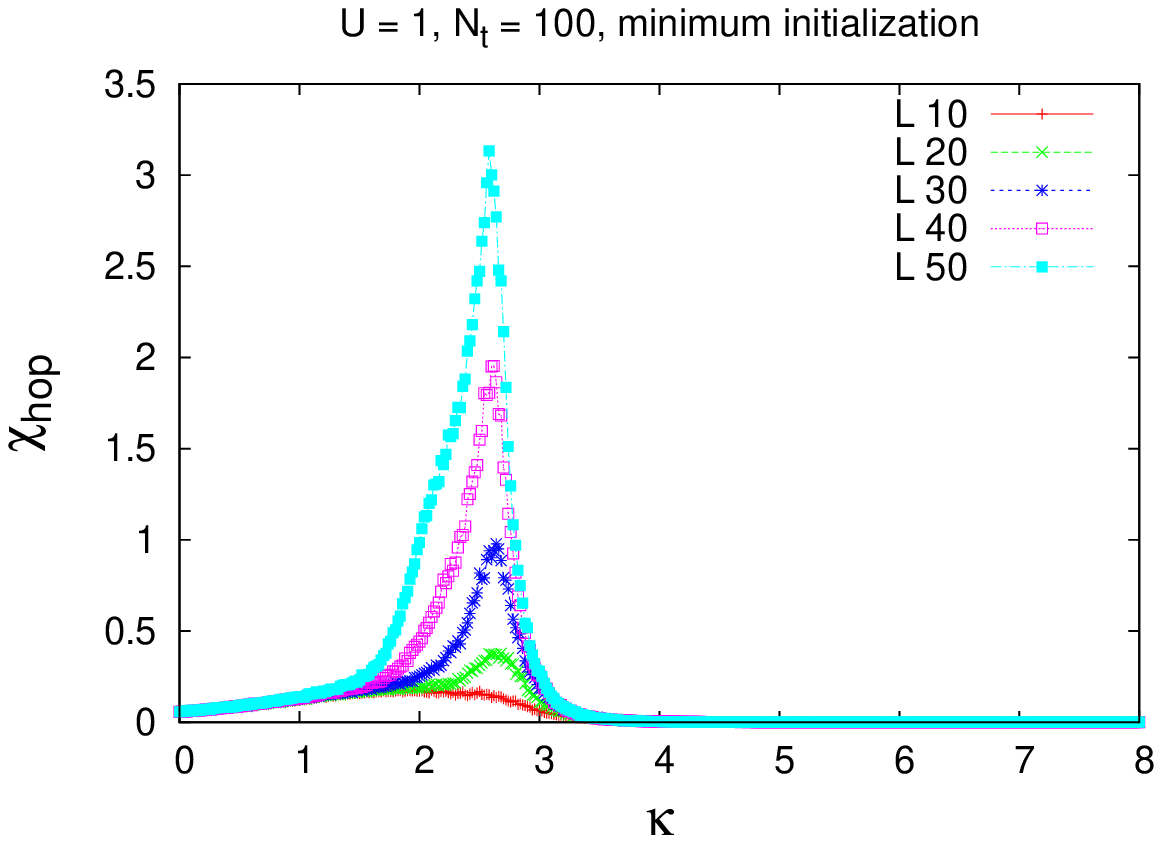}
}
\subfigure[~minimum random]  
{   
 \label{22c}
 \includegraphics[scale=0.5]{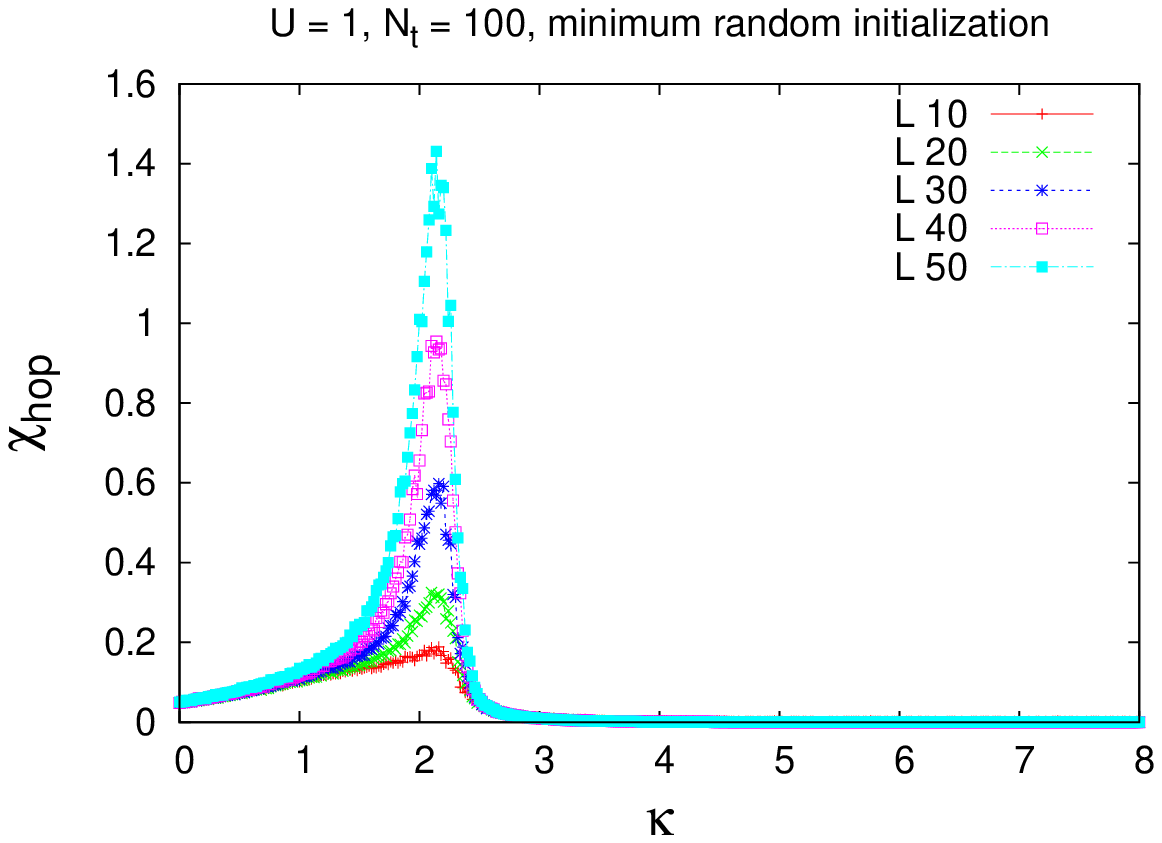}
}
\subfigure[~annealing]  
{   
 \label{22d}
 \includegraphics[scale=0.5]{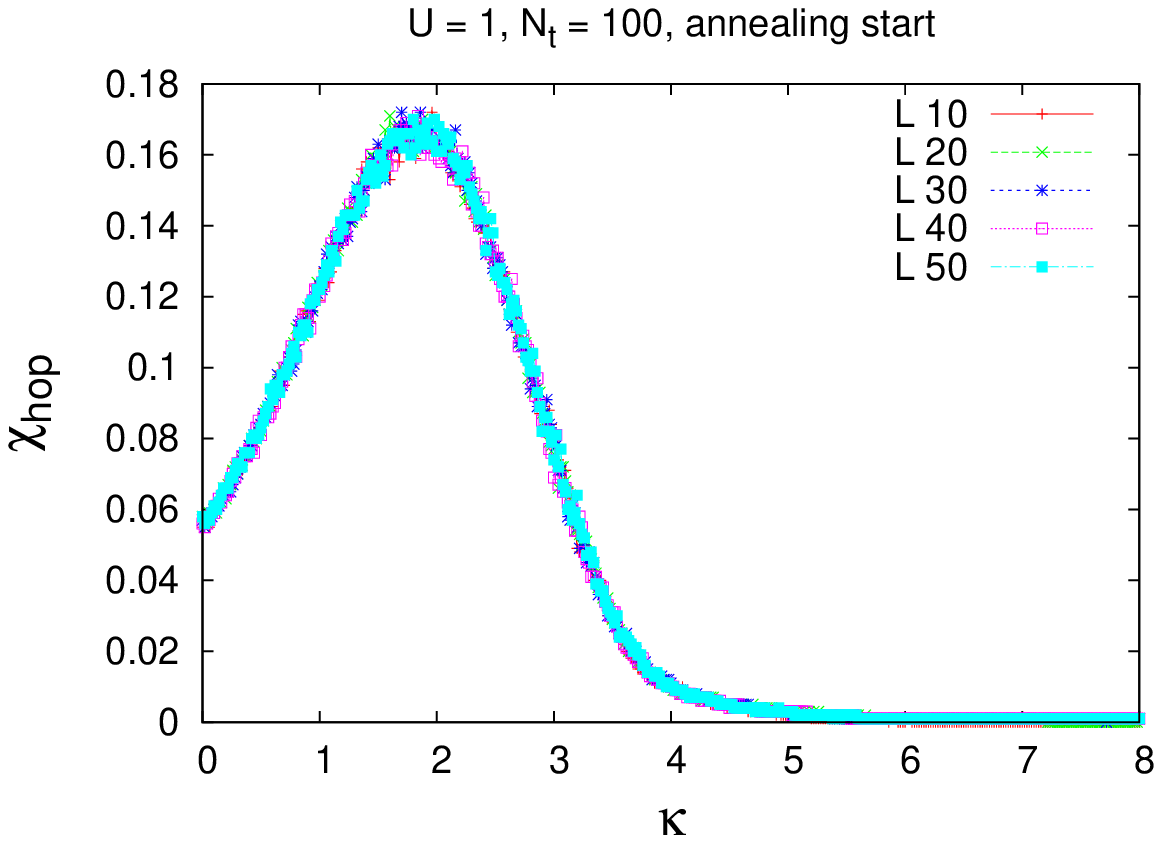}
}
\caption{ $\chi_{hop}$ vs.\ $\k$ at weaker ($U=1$) repulsion and 60\% filling, $N_t=100$.  Initializations:   (a) random;  (b) minimum; (c) minimum random; (d) annealing start configuration.} 
\label{fig22}
\end{figure*}

\begin{figure*}[htbp]
\subfigure[~random]  
{   
 \label{23a}
 \includegraphics[scale=0.5]{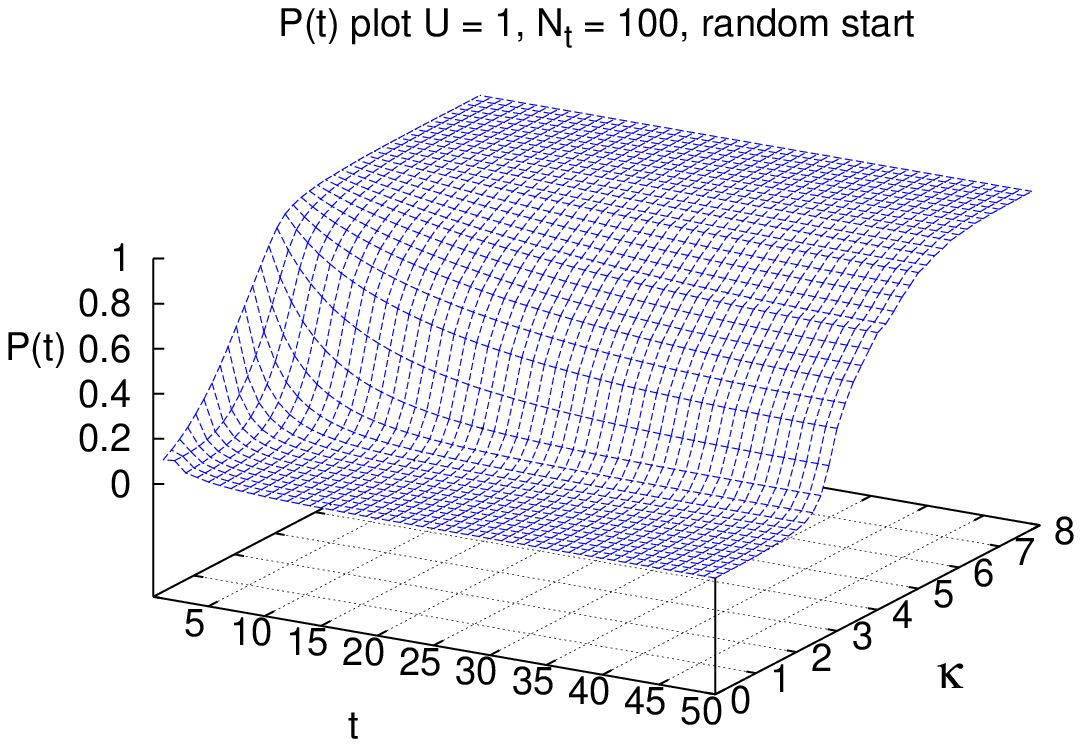}
}
\subfigure[~ minimum]  
{   
 \label{23b}
 \includegraphics[scale=0.5]{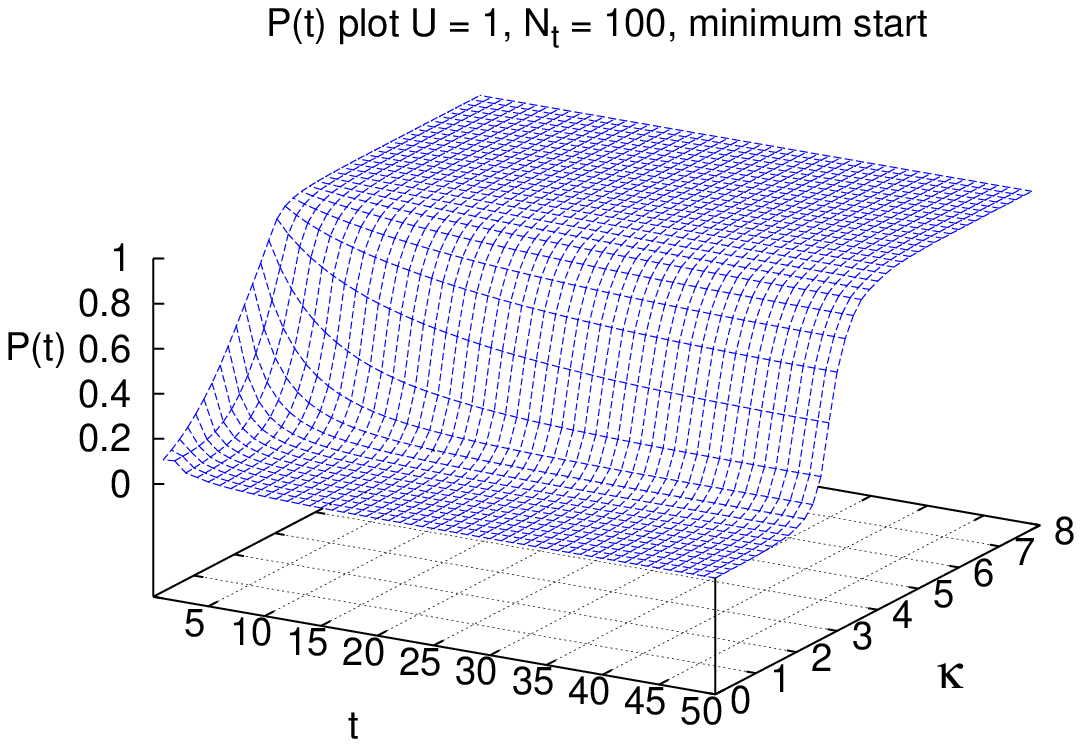}
}
\subfigure[~minimum random]  
{   
 \label{23c}
 \includegraphics[scale=0.5]{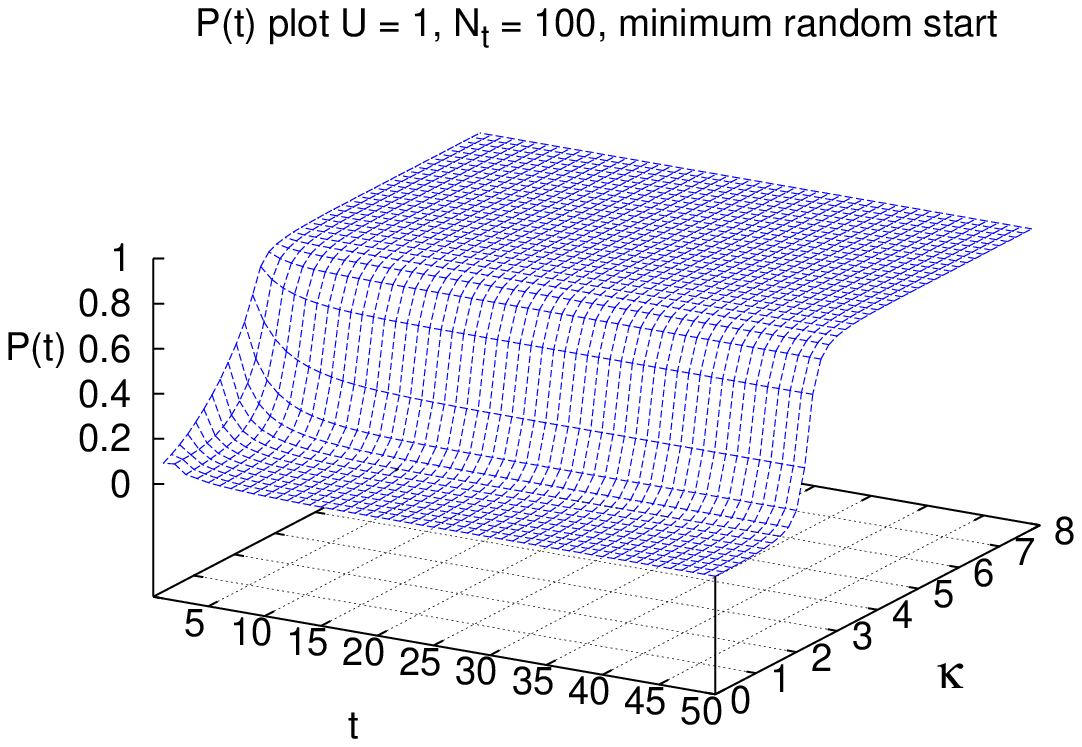}
}
\subfigure[~annealing]  
{   
 \label{23d}
 \includegraphics[scale=0.5]{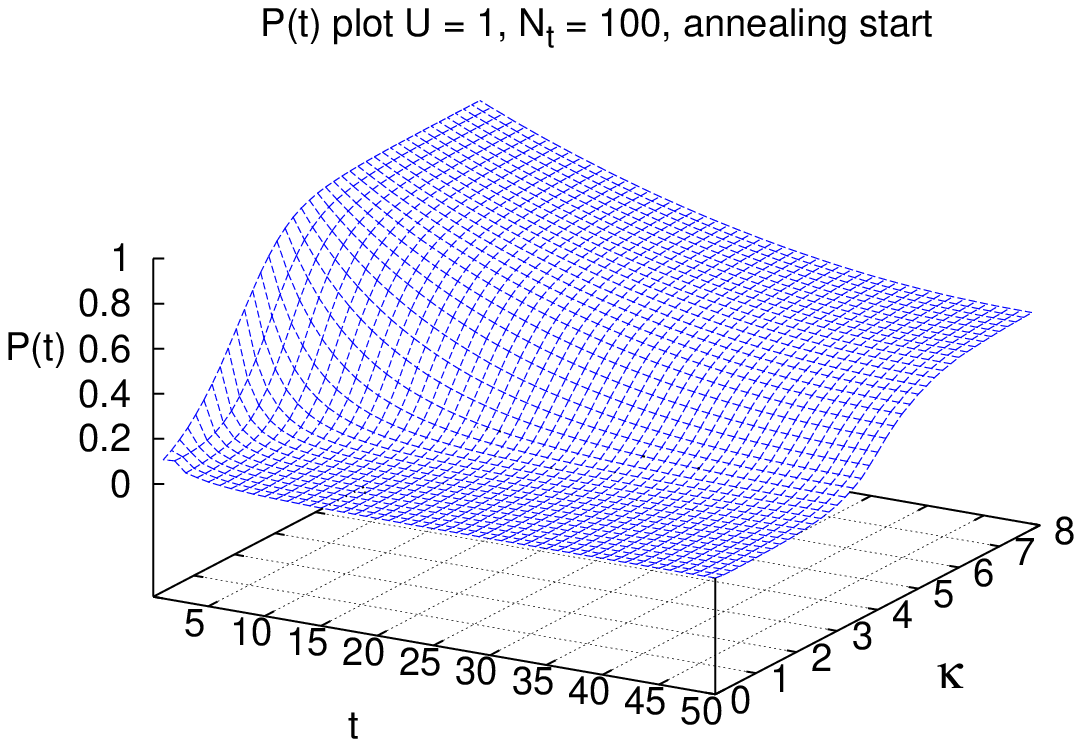}
}
\caption{Localization probability plots at low temperature ($N_t$ = 100) as a function of $\k$ and ~$t$ on a $50\times 50 $ lattice, $U = 1$ and 60$\%$ density. (a) random; 
    (b) minimum; 
(c) minimum random; (d) annealing initial configurations.}
\label{fig23}
\end{figure*}

    The susceptibility and localization data for $U=1$ are shown in Figures \ref{fig22} and \ref{fig23}, and here again
we see features already found at $U=100$, namely the (apparent) quantum phase transition for all but the
annealed initialization, and the fact that the susceptibility data on the smallest $10\times 10$ lattice is not far from the
result for the annealed lattice.  On the other hand, we find that the random, minimum random, and minimum initializations give roughly consistent results (in particular the double peak structure is gone), and it is only annealing initialization that shows no sign of a quantum phase transition.
 
\begin{figure*}[htbp]
\subfigure[~random]  
{   
 \label{32a}
 \includegraphics[scale=0.5]{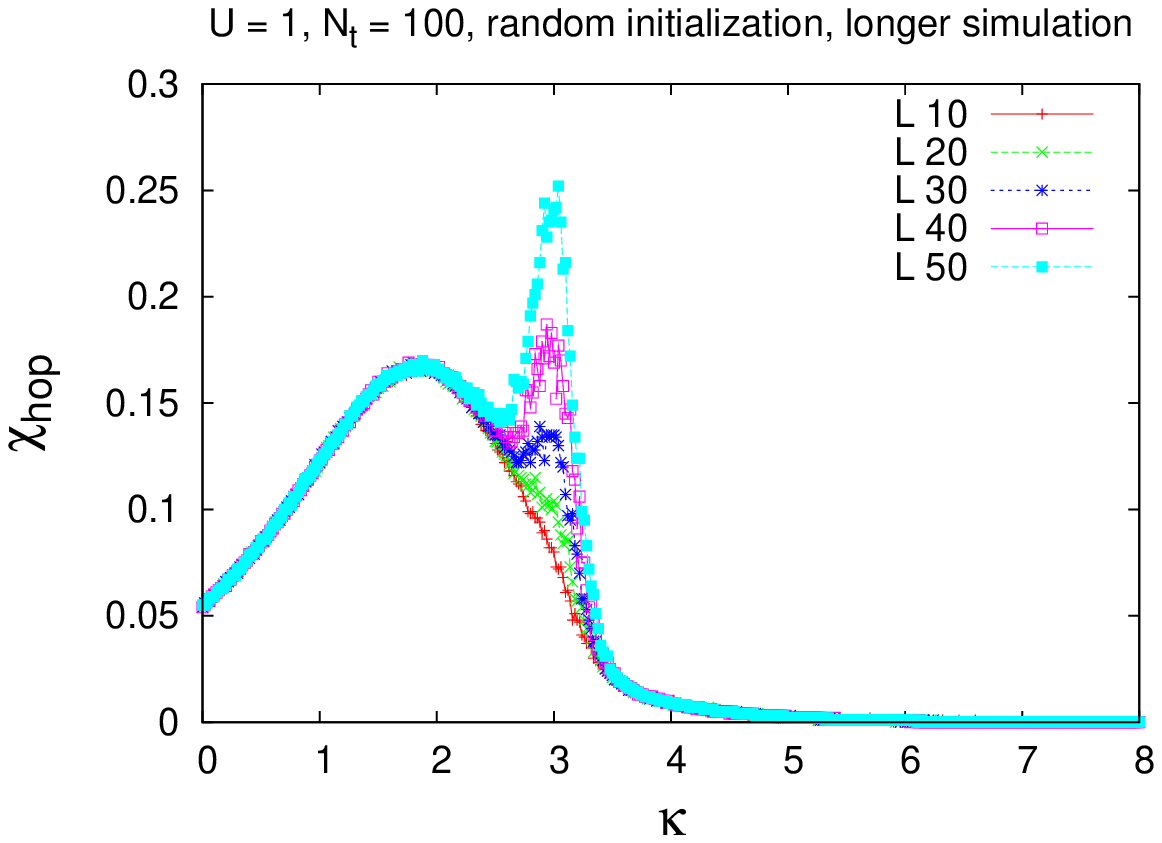}
}
\subfigure[~minimum]  
{   
 \label{32b}
 \includegraphics[scale=0.5]{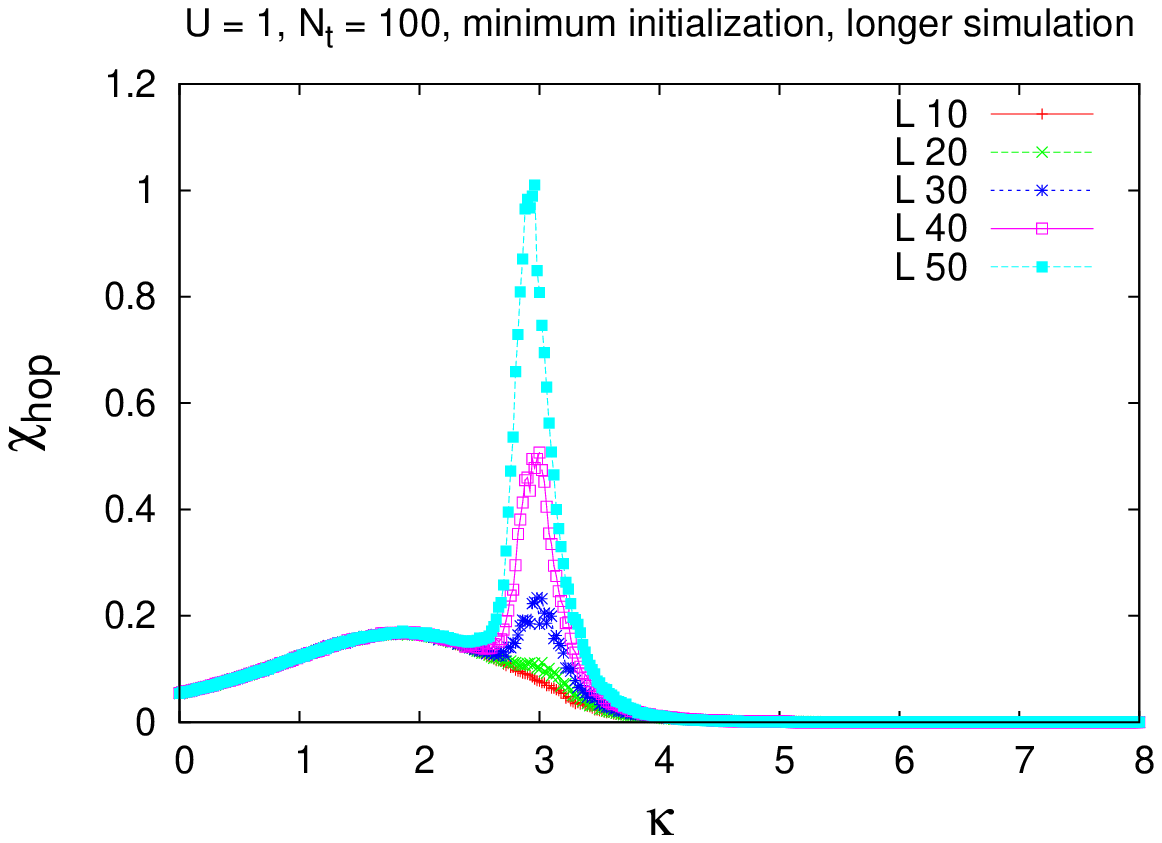}
}
\centering
\subfigure[~minimum random]  
{
\label{32c}
 \includegraphics[scale=0.5]{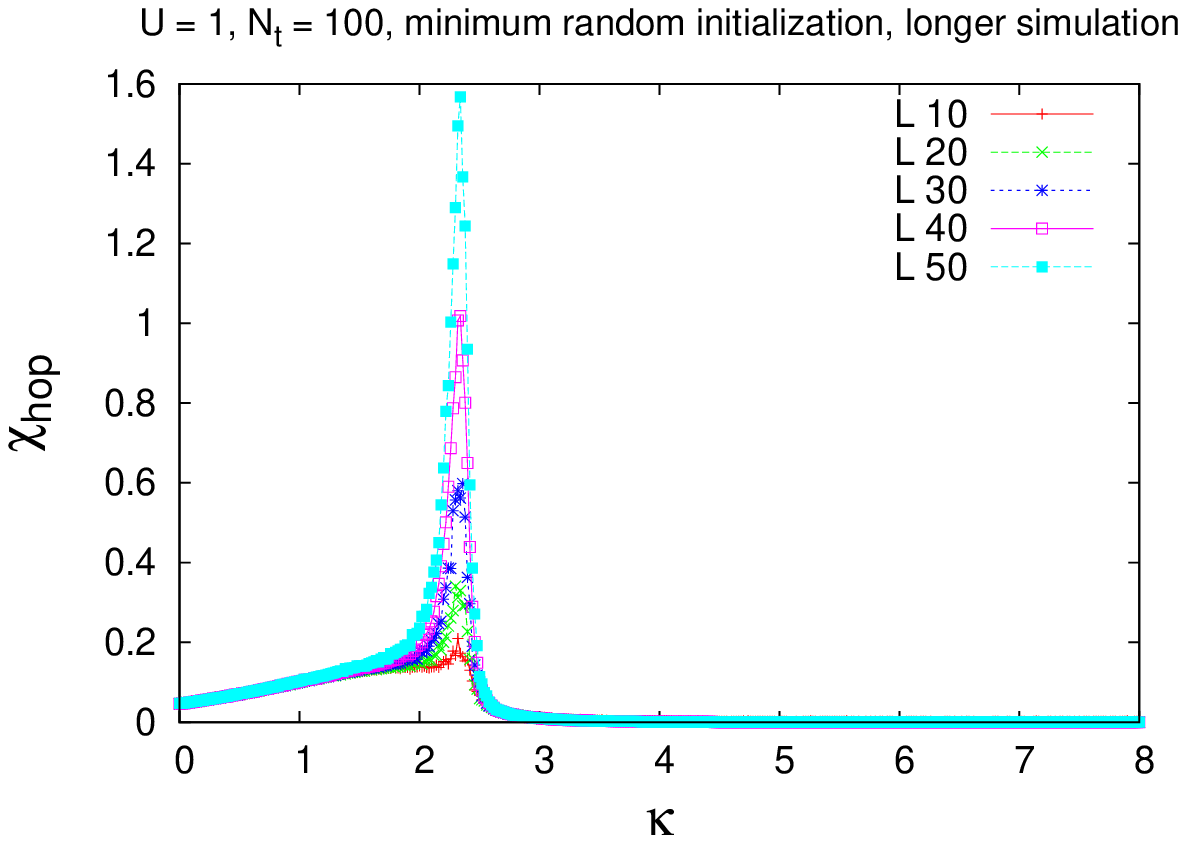}
}
\caption{Same as Figure \ref{fig22}(a-c), with Monte Carlo sweeps increased by a factor of 10, to $10^5$ total. 
Initialization: (a) random; (b) minimum; (c) minimum random starting configurations.} 
\label{fig32}
\end{figure*}

 We also repeated the susceptibility calculations by increasing the number of thermalization and data taking sweeps by an order of magnitude, i.e.\ to $10^5$ total sweeps, in Fig.\ \ref{fig32}.   The random and minimum initial configurations lead to similar results, seen in Fig.\ \ref{32a} and \ref{32b}.  Yet there is still non-ergodicity, because the susceptibility plot corresponding to the minimum random initialization still shows a strong single peak, as high as the one seen in Fig.\ \ref{22c}, which was obtained with an order-of-magniture fewer Monte Carlo sweeps.

\begin{figure*}[htbp]
\subfigure[~random]  
{   
 \label{24a}
 \includegraphics[scale=0.5]{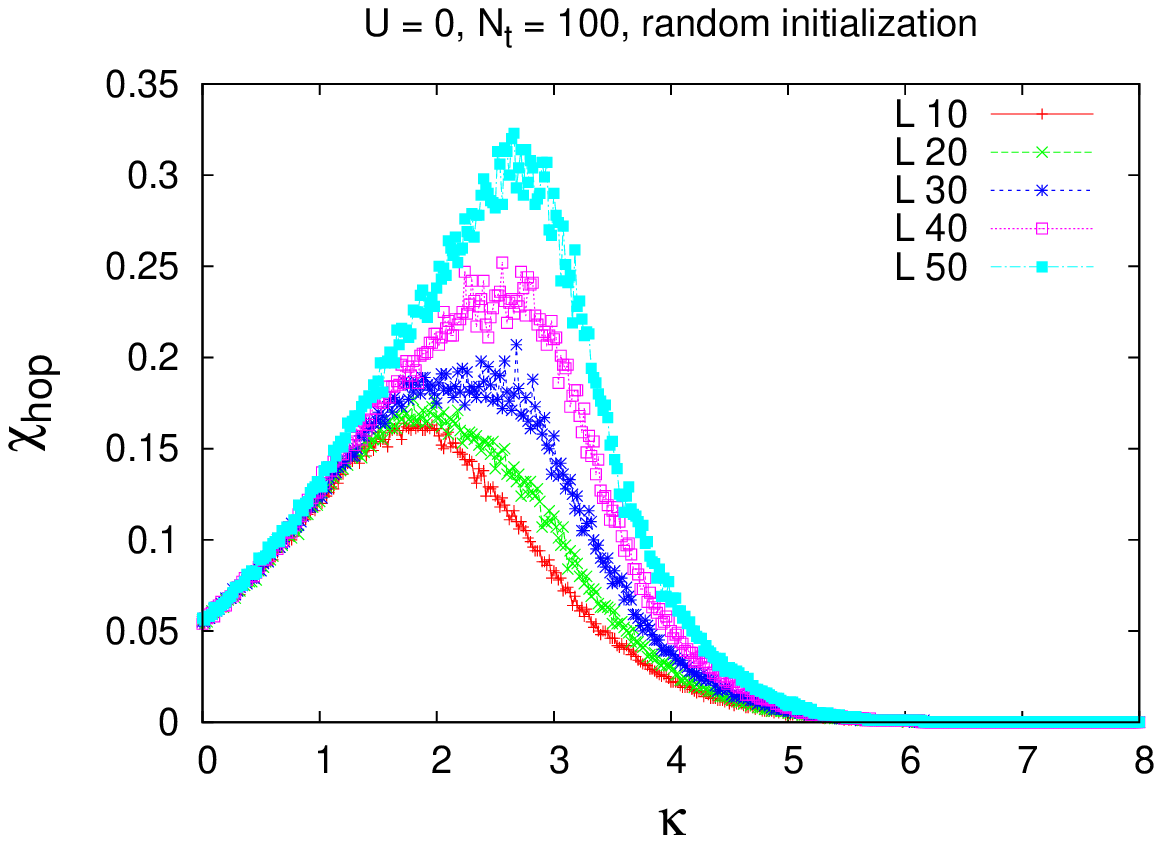}
}
\subfigure[~minimum]  
{   
 \label{24b}
 \includegraphics[scale=0.5]{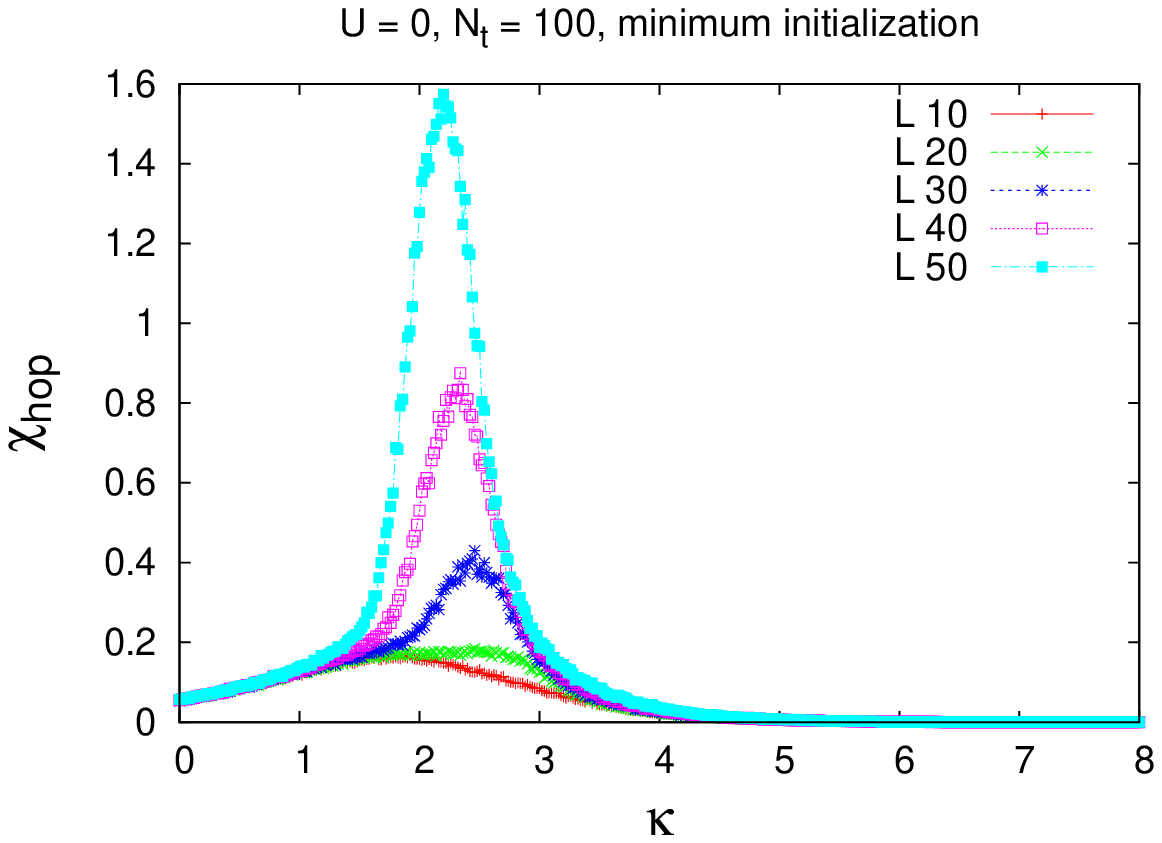}
}
\subfigure[~minimum random]  
{   
 \label{24c}
 \includegraphics[scale=0.5]{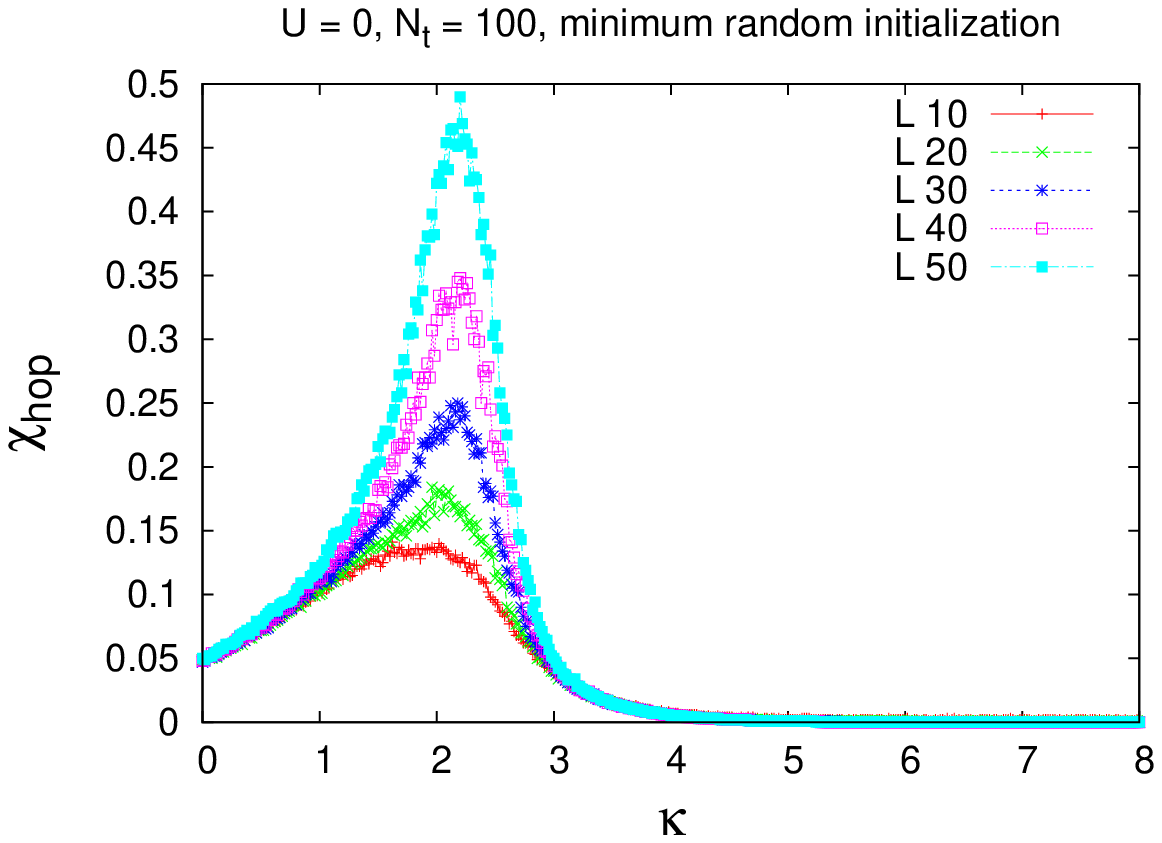}
}
\subfigure[~annealing ]  
{   
 \label{24d}
 \includegraphics[scale=0.5]{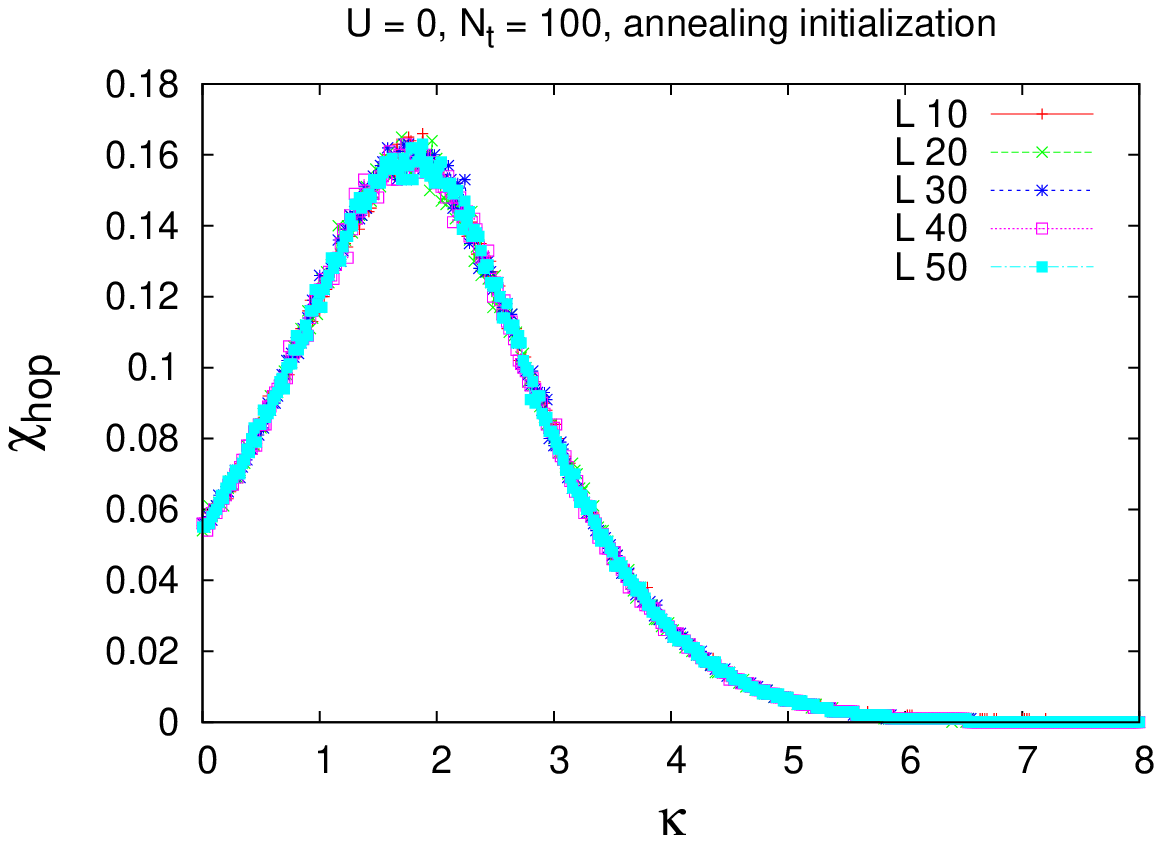}
}
\caption{$\chi_{hop}$ vs.\ $\k$ at zero repulsion ($U=0$), 60\% filling, $N_t=100$.  Initialization (a) random;
(b) minimum; (c) minimum random; (d) annealing initial configurations.} 
\label{fig24}
\end{figure*}

\begin{figure*}[htbp]
\subfigure[~random initial configuration, 10 times longer thermalization time]  
{   
 \label{25a}
 \includegraphics[scale=0.5]{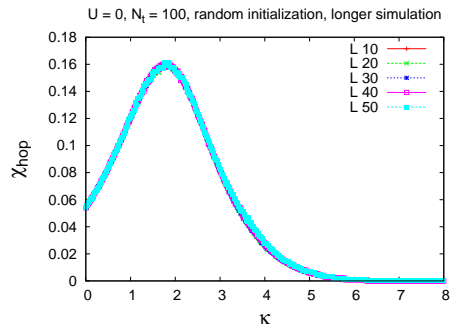}
}
\subfigure[~minimum initial configuration]  
{   
 \label{25b}
 \includegraphics[scale=0.5]{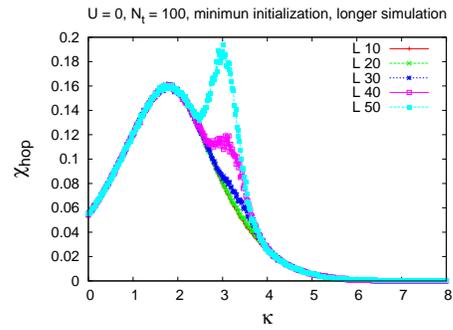}
}
\centering
\subfigure[~minimum random initial configuration]  
{   
 \label{25c}
 \includegraphics[scale=0.5]{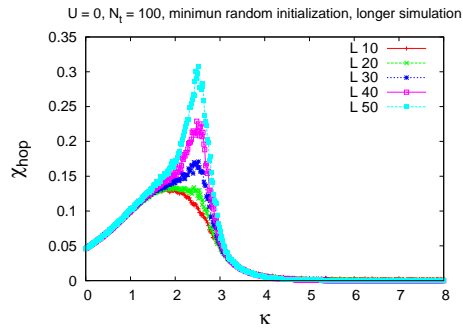}
}
\caption{10 times longer thermalization time. Hopping susceptibility plots of low temperature 
 ($N_t$ = 100) as a function of $\k$, U = 0 of various volume, 60$\%$ filling. (a) random initial configuration.
(b) minimum initial configuration. (c) minimum random initial configuration.} 
\label{fig25}
\end{figure*}
  
In the case of no Coulomb repulsion whatever, i.e.\ $U=0$ and retaining only the Exclusion Principle,
it appears that even when the interaction between the two types of particles is eliminated, the Exclusion Principle is still
sufficient to produce very long relaxation times.  The results for 10,000 total Monte Carlo sweeps, again at 60\% density, 
are shown in Fig.\ \ref{fig24}. Here again there seems to be non-ergodicity, and the random, minimum, and minimum random initializations produce 
results which are very different from the annealed case.  However, if we again increase the Monte Carlo sweeps by a factor of
10, to $10^5$ total sweeps, the situation is different, as shown in Fig.\ \ref{fig25}.  The data for the larger volumes seem to converge, for this larger number of sweeps, to the data which was found on the $10\times 10$ area, which itself fits with the simulated annealing data.  It is true
the larger lattice areas still have some peculiar spikes; we might guess that these will disappear after still longer simulations.
 If so, the comparison of $U=0$ and higher $U$ cases suggests that relaxation times, assuming they are finite at higher $U$,
increase both with lattice size {\it and} with Coulomb repulsion.  Again the analogy to a glass transition is suggestive.

\subsection{Ergodicity at a higher temperature: $N_t = 5, ~ U=100, ~ 60\%$ filling}

\begin{figure*}[htbp]
\subfigure[~random initial configuration at high temperature]  
{   
 \label{27a}
 \includegraphics[scale=0.5]{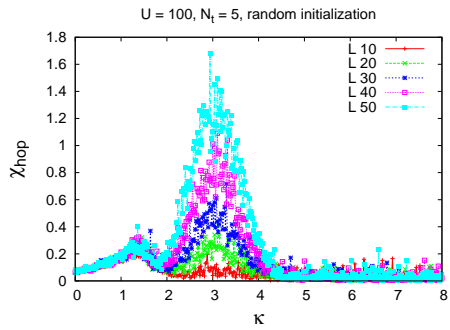}
}
\subfigure[~minimum initial configuration at high temperature]  
{   
 \label{27b}
 \includegraphics[scale=0.5]{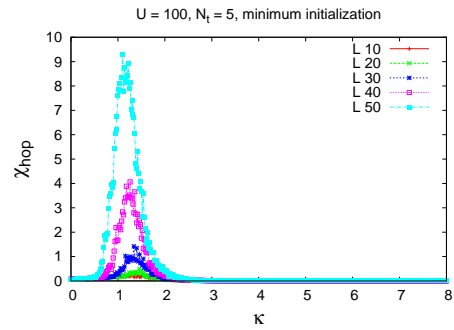}
}
\subfigure[~minimum random initial configuration at high temperature]  
{   
 \label{27c}
 \includegraphics[scale=0.5]{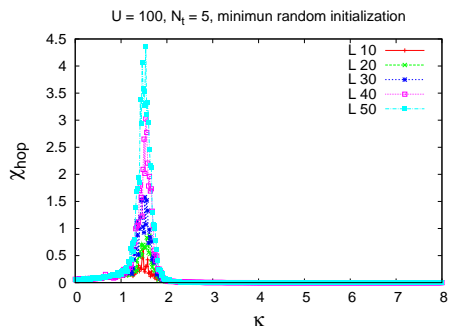}
}
\subfigure[~annealing at high temperature]  
{   
 \label{27d}
 \includegraphics[scale=0.5]{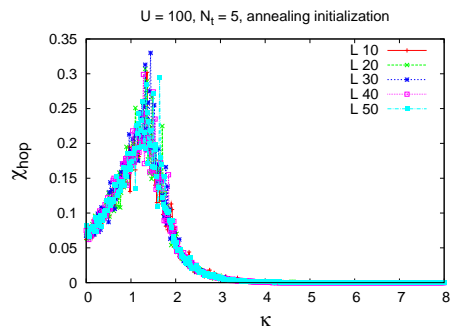}
}
\caption{Effect of increasing temperature by a factor of 20, to $N_t=5$.  As in Figure \ref{fig456} we set the density at 60\% with strong repulsion ($U=100$).} 
\label{fig27}
\end{figure*}

The next question is whether ergodicity is restored (or relaxation times reduced) at high temperature.
In this connection we decrease the time extension from $N_t=100$ to $N_t=5$, thereby increasing the temperature by a factor of 20,  and  again calculate the hopping susceptibility for various different initial particle configurations at $U=100$ and 60\% filling.  The results are shown in Fig.\ \ref{fig27}.  All of susceptibility plotting results are different from each other, indicating as before lack of ergodicity, at least up to
10000 total sweeps.

The one difference we observe here is in random initial configuration result in Fig.\ \ref{27a}, there the first peak existing in the low temperature plot in Fig.\ \ref{fig4} disappeared in high temperature plot. The single peaks arising in the minimum and minimum random initializations at low temperature remained in the high temperature plots. The positions of those single peaks at the higher and lower temperatures are found about the same $\k$ values.  With the annealing initialization, the volume dependent peaks do not exist in Fig.\ \ref{27d}.

 \begin{figure*}[htbp]
\subfigure[~random]  
{   
 \label{28a}
 \includegraphics[scale=0.5]{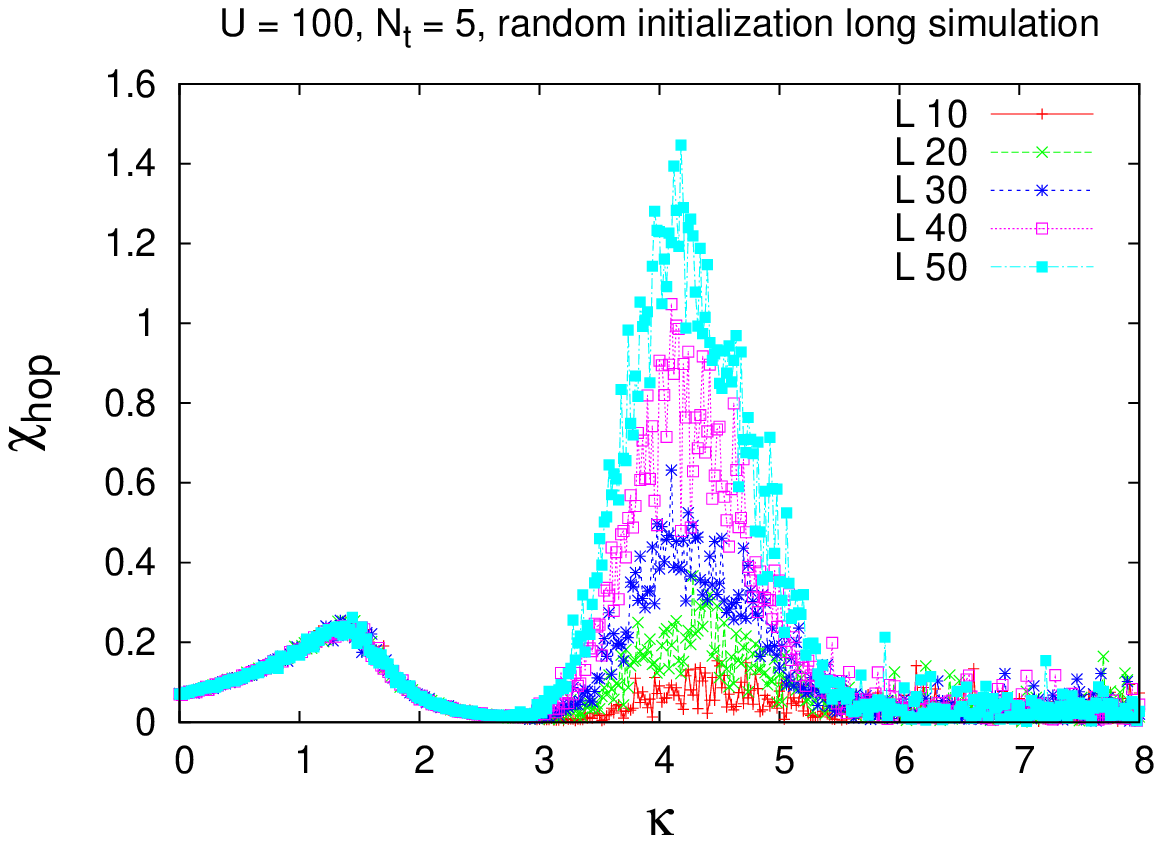}
}
\subfigure[~minimum initial configuration]  
{   
 \label{28b}
 \includegraphics[scale=0.5]{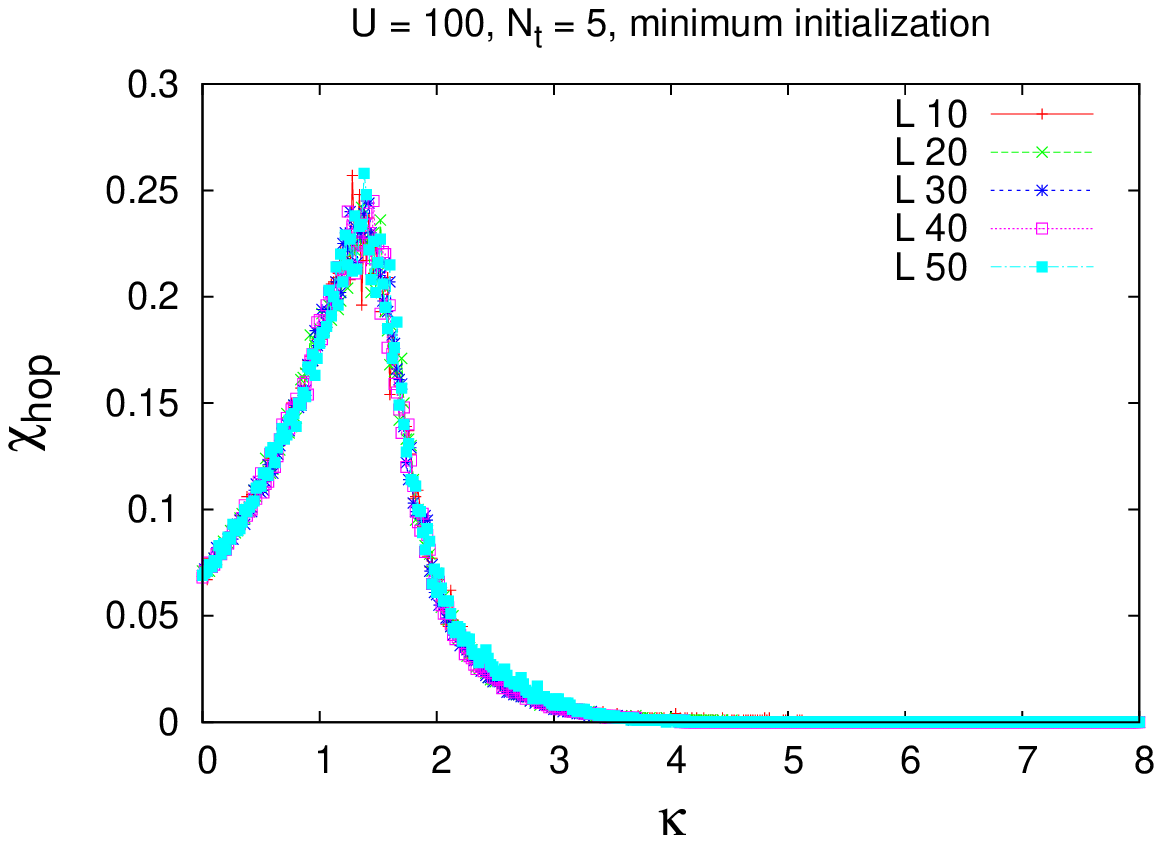}
}
\centering
\subfigure[~minimum random initial configuration]  
{   
 \label{28c}
 \includegraphics[scale=0.5]{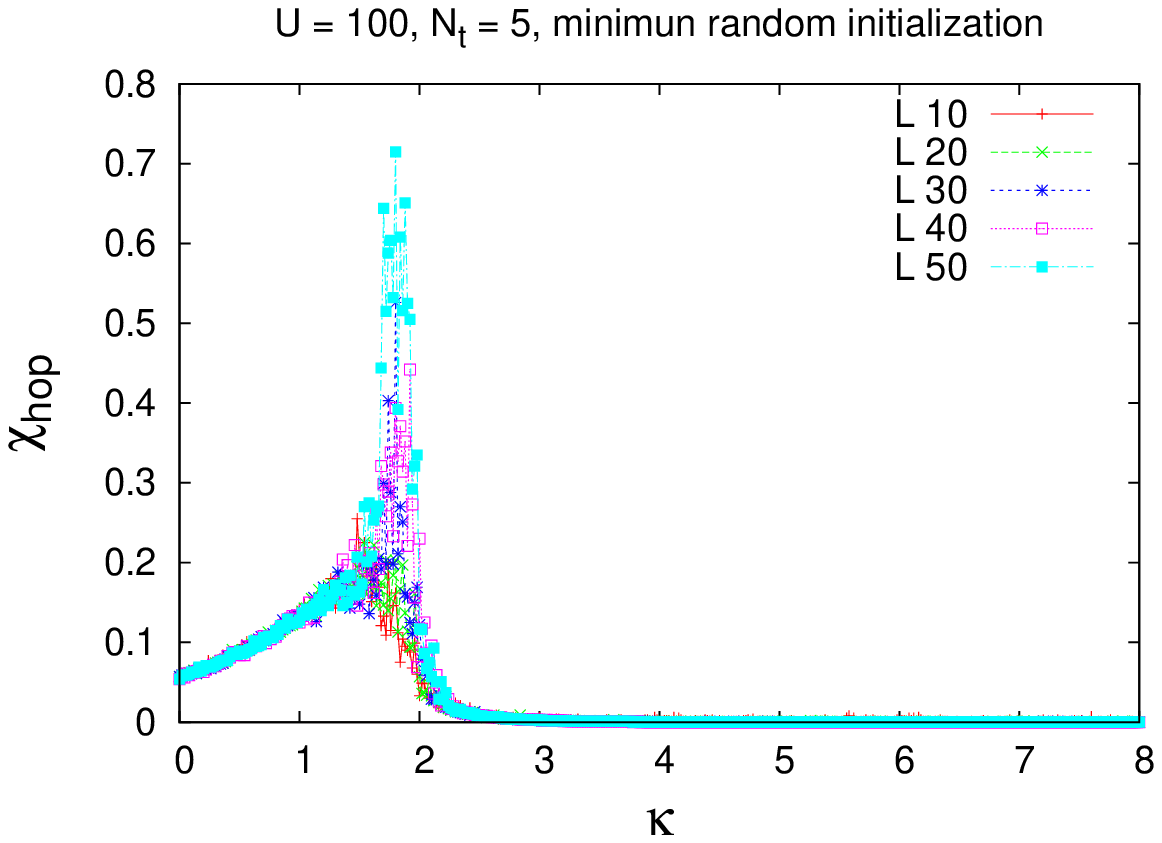}
}
\caption{Same as Figure \ref{fig27} at high temperature, with Monte Carlo sweeps increased by an order of magnitude, to $10^5$ total sweeps.
Initializations: (a) random; (b) minimum; (c) minimum random initial configuration.} 
\label{fig28}
\end{figure*}
 
Now we again increase both the thermalization and data taking sweeps by a factor of 10, to $10^5$ total update sweeps, and repeat our calculations in Fig.\ \ref{fig28}.   We find that this longer simulation time eliminates the volume dependent single peaks in the hopping susceptibility plot in minimum and minimum random configurations in Fig.\ \ref{28b} and Fig.\ \ref{28c}, and brings both of these plots close to the annealed result in Fig.\ \ref{27d}.\footnote{There is still a spike in the minimum random data, but its height is greatly reduced compared to Fig.\ \ref{27c}, but we may guess that this spike will disappear entirely in a still longer simulation.}  Yet even in this case, the data for random initialization has not converged to the annealed start; in that the second peak remains.  Its position has shifted to a larger $\k$ value, as compared to the fewer sweeps result in Fig.\ \ref{27a}, but the height of the second peak has stayed about the same.

So the data regarding ergodicity at this much higher temperature are not unambiguous.  The annealed, minimum, and minimum random initializations tend, eventually, to the same result.  But the random initialization, with the peculiar second peak, still differs, and the question is whether this difference is significant.   

\subsection{\label{peak2} The second peak at random initialization}
\begin{figure*}[htbp]
\subfigure[~low temperature]  
{   
 \label{29a}
 \includegraphics[scale=0.5]{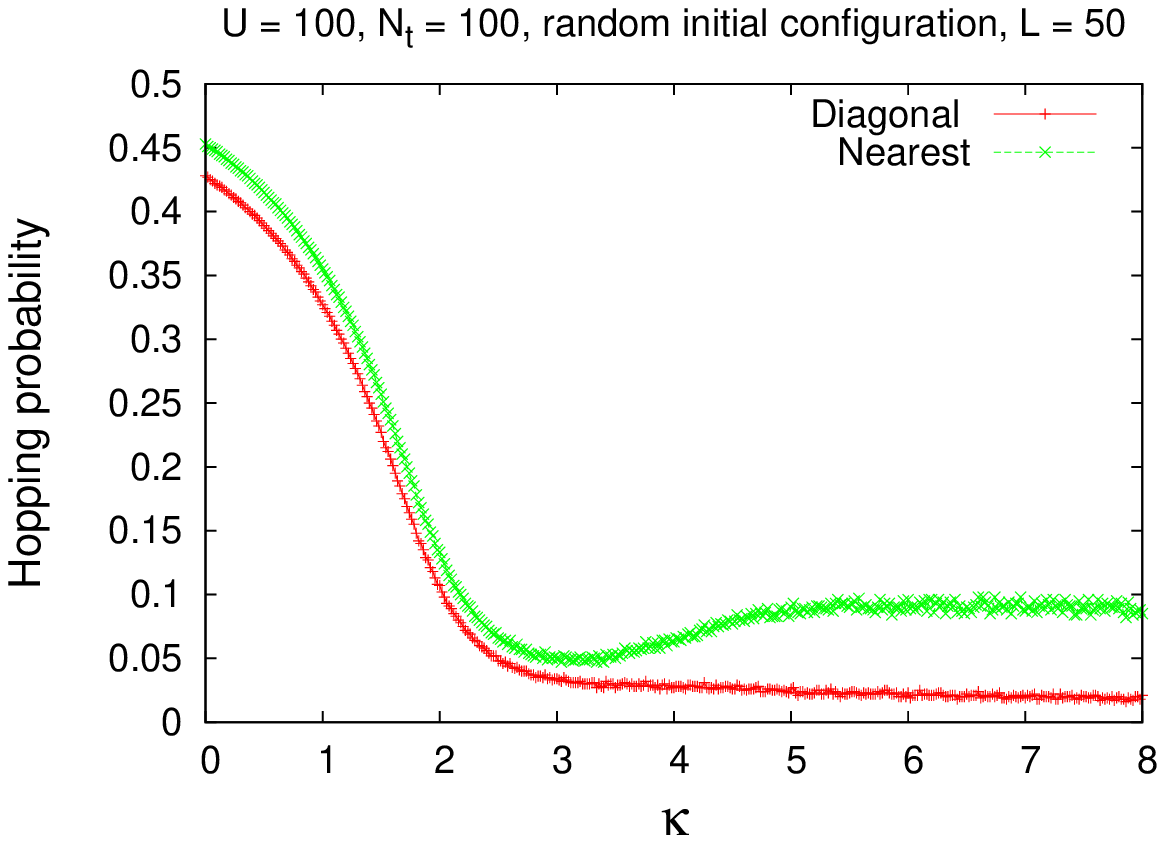}
}
\subfigure[~high temperature]  
{   
 \label{29b}
 \includegraphics[scale=0.5]{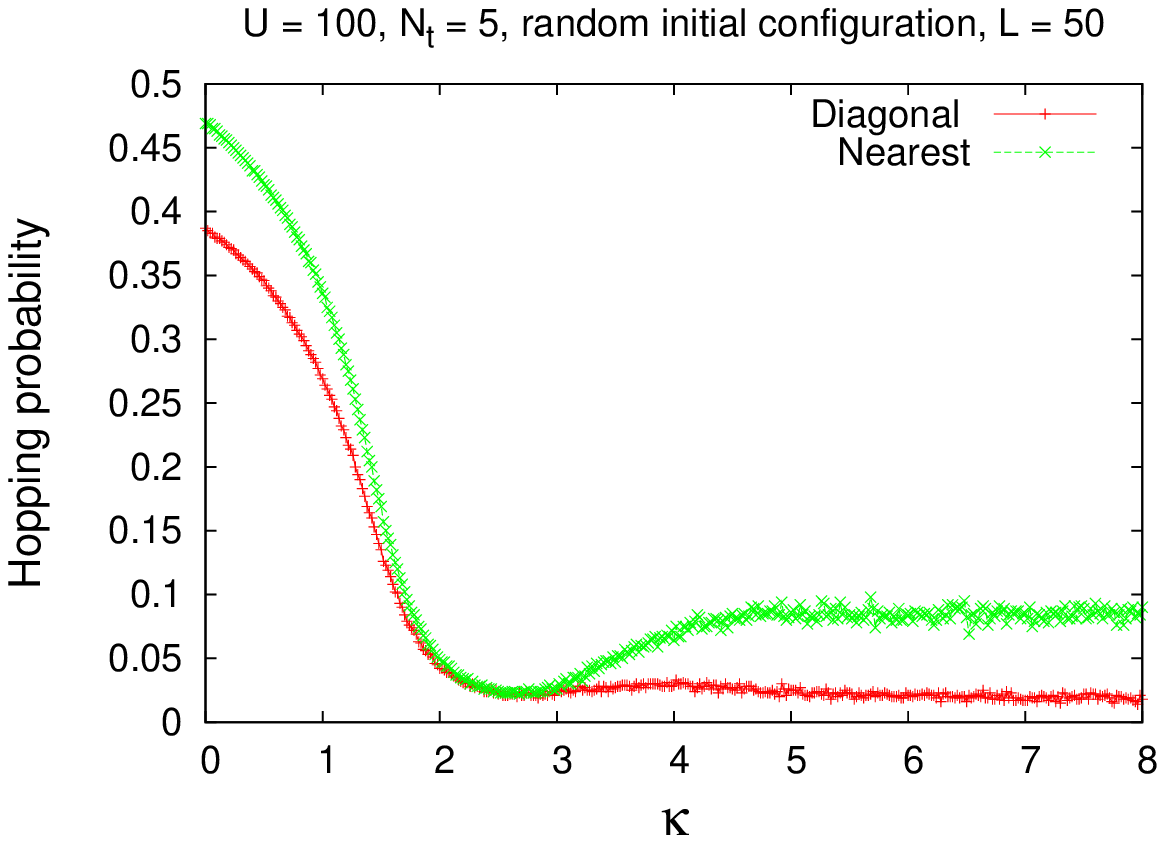}
}
\caption{Diagonal and nearest-neighbor hopping probabilities, as defined in eqs.\ \rf{dhop} and \rf{nhop}, at $U=100$, 60\% filling, random initialization, and  
    (a) low temperature, $N_t$ = 100 (b) high temperature, $N_t$ = 5.} 
\label{fig29}
\end{figure*}

To study the nature of the second peak, we have computed the average hopping probability for nearest neighbor and next-nearest neighbor (diagonal) separately, with hopping contributions denoted $j_+$ and $j_\times$ respectively, corresponding to a hop at time $t$ to a nearest or next nearest site at time $t+1$. We work again at $U = 100$ and 60\% filling at the higher and lower temperatures of $N_t=5$ and $N_t=100$, and define the observables on a time-slice (see Fig.\ \ref{fig29}).

  \beq
          \text{NN hop}(t) = {1\over n_p} \sum_{n=1}^{n_p} \oh( j_+(n,t-1) + j_+(n,t)) 
         \label{nhop}
  \eeq
  \beq
         \text{Diagonal hop}(t) = {1\over n_p} \sum_{n=1}^{n_p} \oh( j_\times(n,t-1) + j_\times(n,t)) 
         \label{dhop}
 \eeq
 
    What is the explanation of this strange second peak which appears with the random initialization, and of the hopping probability which remains surprisingly high at large values of $\k$?  And why is this peak, and this surprisingly high hopping probability at large 
$\k$ not also seen for the minimum and minimum random initializations?  We think the reason is actually straightforward, at least as concerns the hopping probability.  With a random initialization, in contrast to the minimum and minimum random configurations, the system begins the simulation with some number of unoccupied sites.   At $U=100$ there is a very strong tendency for a particle in a double-occupied site to move to a neighboring empty site; the cost in $K(n)$ is more than made up for by the drop in the very high potential energy.  Of course this introduces some degree of hopping along many of the particle trajectories, even at rather large $\k$.  At some point there are no more unoccupied sites that can be
accessed by a single hop from double occupied sites, and the trajectories are at that stage ``frozen''  at large $\k$.  This is because any further hopping, even from a double occupied site to a single occupied site (which costs nothing in potential energy) is strongly disfavored by the kinetic part of the action.\footnote{Fluctuations which do not alter either the kinetic or potential energies are still permissable.}  However, the hopping probabilities in these frozen trajectories are still non-negligible.  By comparison, in the minimum and minimum random configurations there are no unoccupied sites in the initial configuration, and all hopping is strongly disfavored at large
$\k$ values. 

    The conclusion is that the hopping probabilities for random initialization at large $\k$ originate from particles moving towards a lower potential energy configuration in the early part of the simulation, and the hopping which is essentially frozen into particle trajectories is almost entirely due to the motion at that period.  The second peak at high density and random initialization may perhaps be associated with a transition to a phase of ``frozen'' trajectories which still exhibit substantial hopping in the time direction along some trajectory.  

\subsection{Lower density}

\begin{figure*}[htbp]
\subfigure[~random]  
{   
 \label{33a}
 \includegraphics[scale=0.5]{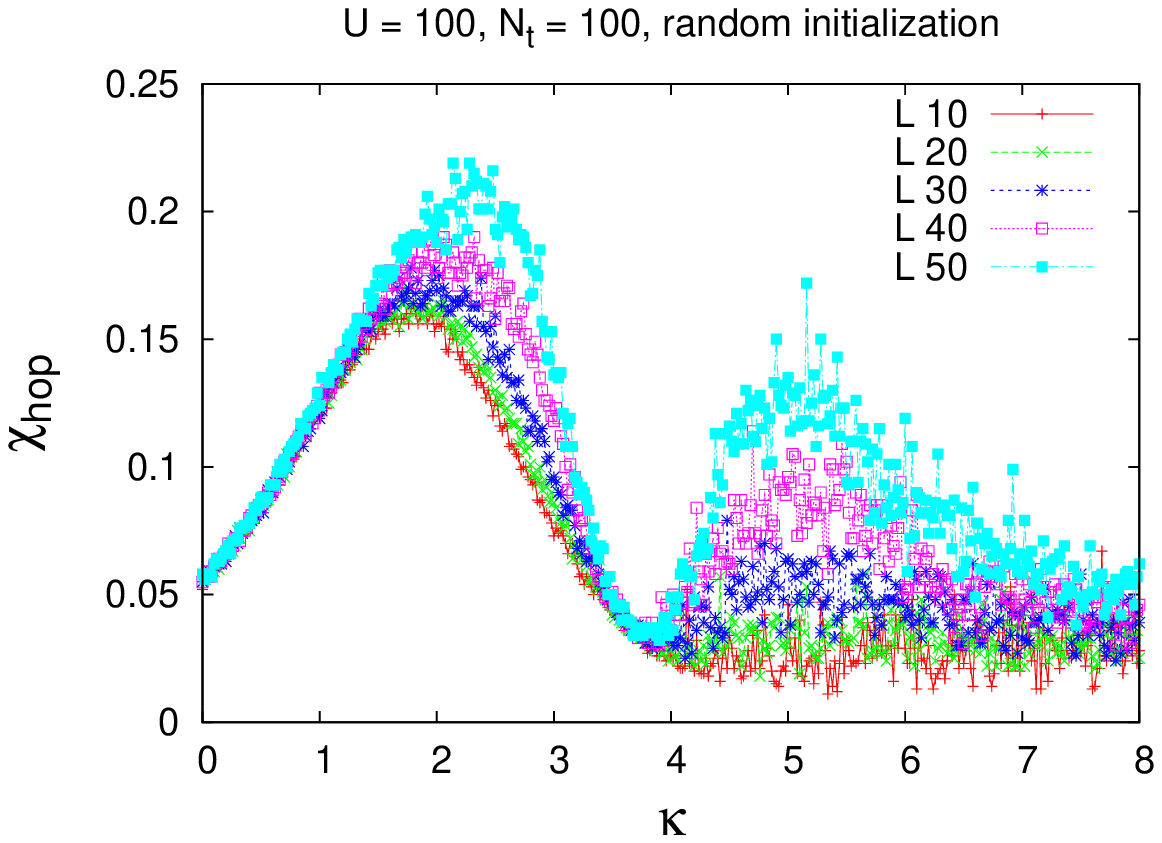}
}
\subfigure[~random double]  
{   
 \label{33b}
 \includegraphics[scale=0.5]{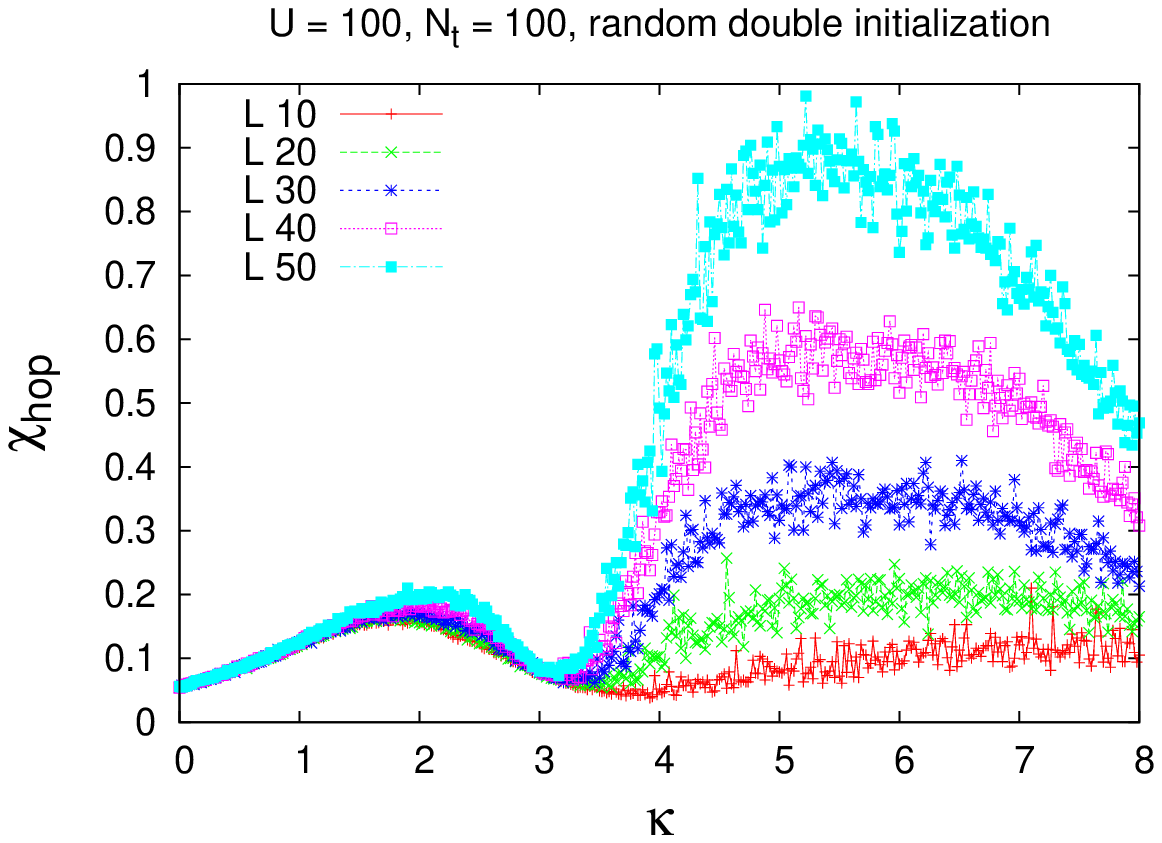}
}
\subfigure[~annealing]  
{   
 \label{33c}
 \includegraphics[scale=0.5]{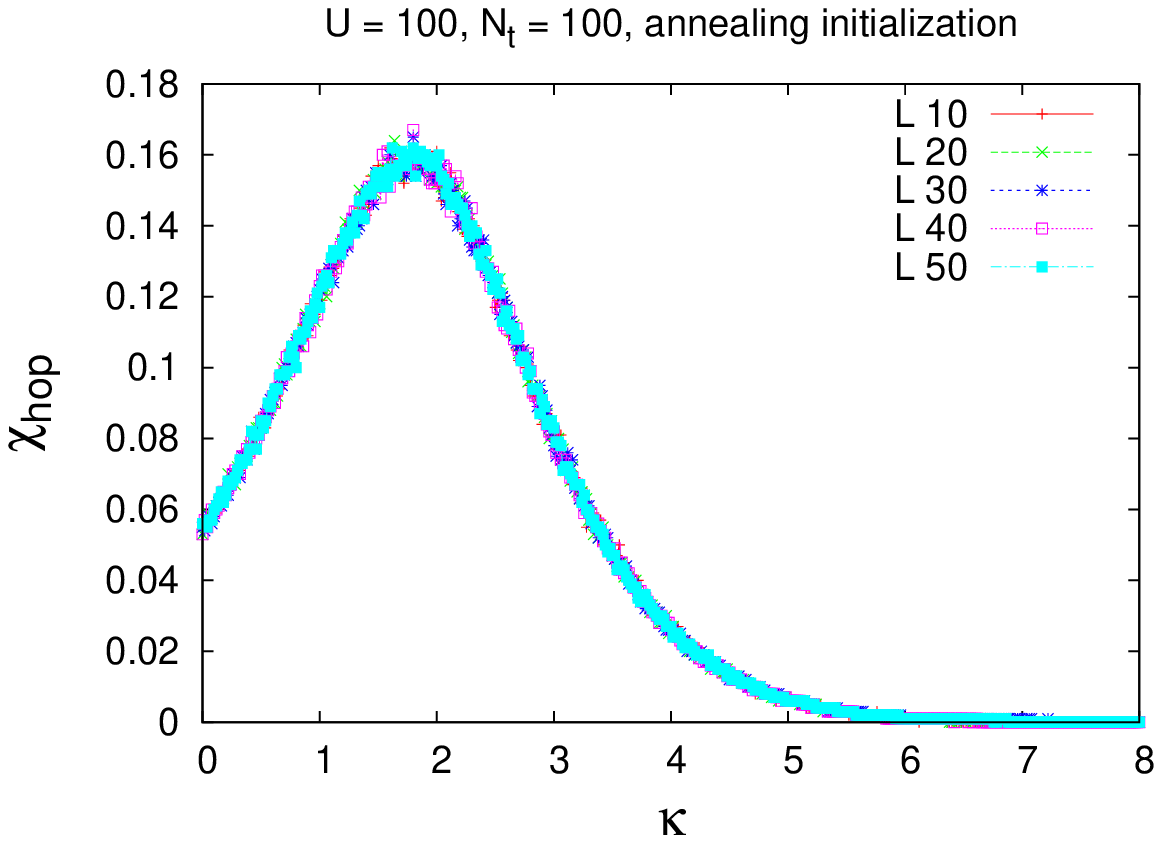}
}
\subfigure[~random, $10^5$ MC sweeps]  
{   
 \label{33d}
 \includegraphics[scale=0.5]{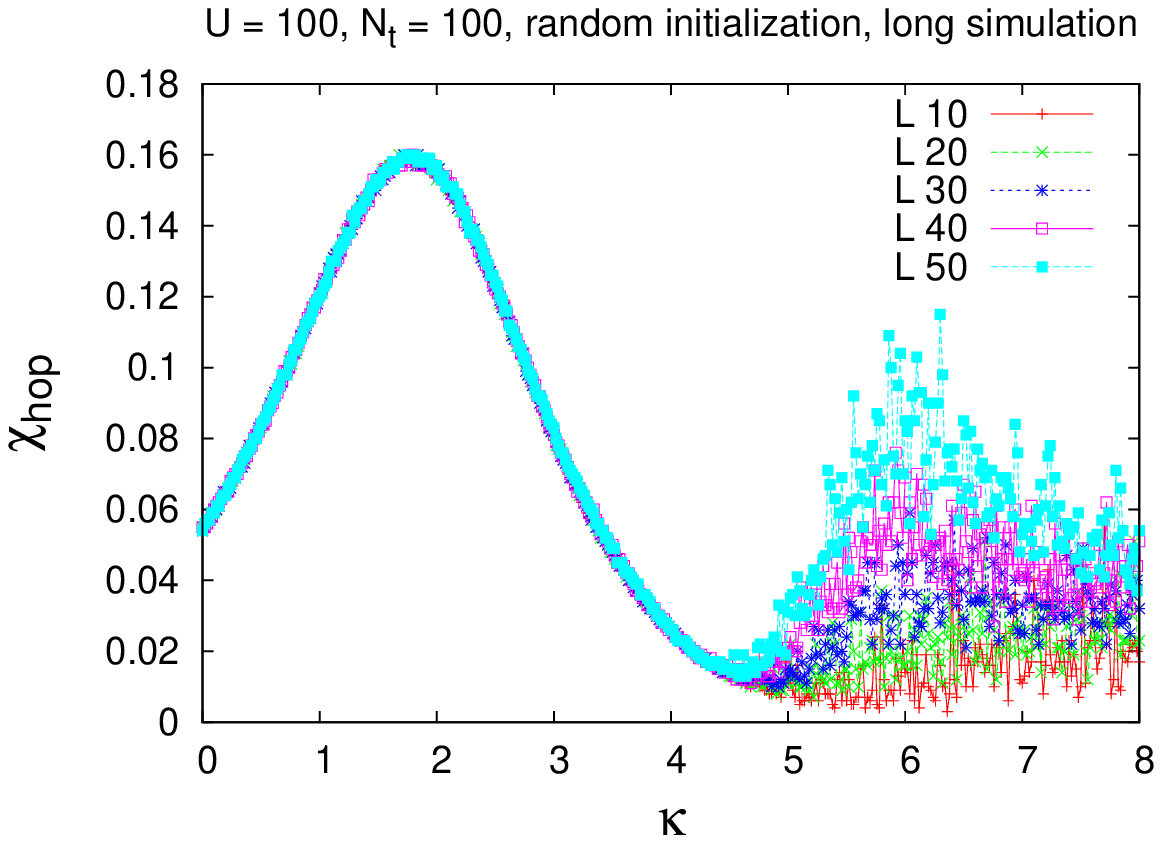}
}
\caption{Lower density, 30\% filling.  $\chi_{hop}$ vs. $\k$ at strong repulsion ($U=100$) and low temperature ($N_t=100$).
Initializations: (a) random; (b) random double; (c) annealing initial configurations.  Subfigure (d) is for a random initial configuration, with Monte Carlo sweeps increased by a factor of 10.} 
\label{fig33}
\end{figure*}

    We would expect that the non-ergodicity that we have observed at 50\% and 60\% filling is a high density phenomena,
which one would not see at much lower densities.  As a check, we have repeated our calculations at 30\% filling at
$U=100,~N_t=100$ for several different initializations.  In this situation, the initializations we have described as ``minimum''
and ``minimum random'' do not apply, because there are  simply not enough particles to have at least one particle at every
site.   We introduce instead a new initial configuration where we start with sites which are either unoccupied or double
occupied, with the double occupied sites chosen randomly on the lattice.  We refer to this initialization as ``random double.'' As before, there is no variation in time in the initial random double configuration.  The results are shown in Fig.\  \ref{fig33}.  
Of course there are the striking double peaks at large $\k$ which, as just explained, are a phenomenon from the initial period of the simulation in the random and random double starts. Looking aside from the second peak, at lower $\k$, the random, random double, and annealed initializations
are quite similar, with the relatively small volume dependence seen in Fig.\ \ref{33a} no longer visible when the Monte Carlo
sweeps are increased by a factor of 10, in Fig.\ \ref{33d}.  There are no indications of a quantum phase transition.  
We see, as expected, ergodicity at lower density and moderate values of $\kappa$, while fluctuations are essentially frozen (this is the origin of the second peak) at large $\kappa$.

 \section{Conclusions}

The correspondence between the quantum mechanics of point-like particles, and the statistical mechanics of line-like particle trajectories, suggests a possible source of non-ergodicity, if a dense system of line-like trajectories  begins to exhibit  ``glassy''
behavior, where the system becomes stuck in a rather localized region in  the space of relevant configurations.  
In this article we have investigated the possible loss of ergodicity at high densities in a simple hopping model, inspired
by certain features of the Hubbard model, in which ``spin up'' and ``spin down'' particles can hop in Euclidean time on a two
dimensional lattice.  We have seen that the behavior of this model depends very strongly on the initial starting configuration of the Monte Carlo simulation, and in fact we have seen apparent quantum phase transitions for some starting configurations, but not
for others.  This dependence on initialization occurs at high densities, and we have illustrated the situation at 50\% and 60\%
of maximum filling.  We see that ergodicity is not an issue at lower densities.  Whether this non-ergodic behavior at high densities might be relevant in some experimental situations, e.g.\ systems of ultra cold atoms, remains to be seen. \\

\ni{\bf Acknowledgments}\\

I thank Jeff Greensite for helpful discussions.
  
 \bibliography{shm.bib}
 
\end{document}